\documentclass[runninghead]{llncs}
\pdfoutput=1
\usepackage{fancyhdr} 
\usepackage{caption}
\usepackage{subcaption}
\usepackage{makecell}

\usepackage[most]{tcolorbox}
\tcbuselibrary{listings,breakable,}

\usepackage[ruled]{algorithm2e}
\usepackage{perpage}
\MakePerPage{footnote}

\usepackage{wrapfig}

\usepackage{booktabs,tabulary,array,parskip}
\usepackage{xcolor}

\usepackage[flushleft]{threeparttable}
\usepackage{makecell}
\usepackage{comment}

\usepackage{graphicx}
\usepackage{subfiles}
\usepackage{float}

\usepackage{mathtools}

\usepackage{lscape}
\usepackage[figuresright]{rotating}
\usepackage[graphicx]{realboxes}
\usepackage{adjustbox}
\usepackage{listings}
\usepackage{booktabs}
\usepackage{multirow}

\usepackage{tikz}
\usetikzlibrary{arrows.meta}
\usepackage{framed} 

\usepackage{amsmath}
\usepackage{amssymb}

\usepackage{lscape}
\usepackage{appendix}

\usepackage{comment}

\usepackage{setspace}
\usepackage{ntheorem}

\newtheorem{defi}{Definition}

\usepackage{hyperref} 
\makeatletter
\def\UrlAlphabet{%
      \do\a\do\b\do\c\do\d\do\e\do\f\do\g\do\h\do\i\do\j%
      \do\k\do\l\do\m\do\n\do\o\do\p\do\q\do\r\do\s\do\t%
      \do\u\do\v\do\w\do\x\do\y\do\z\do\A\do\B\do\C\do\D%
      \do\E\do\F\do\G\do\H\do\I\do\J\do\K\do\L\do\M\do\N%
      \do\O\do\P\do\Q\do\R\do\S\do\T\do\U\do\V\do\W\do\X%
      \do\Y\do\Z}
\def\UrlDigits{\do\1\do\2\do\3\do\4\do\5\do\6\do\7\do\8\do\9\do\0}
\g@addto@macro{\UrlBreaks}{\UrlOrds}
\g@addto@macro{\UrlBreaks}{\UrlAlphabet}
\g@addto@macro{\UrlBreaks}{\UrlDigits}
\makeatother

\usepackage{dashbox}
\usepackage{tikz}
\usepackage{enumerate}
\usepackage{enumitem}

\usepackage{etoolbox}
\usepackage{tkz-euclide}

\usepackage{tikz}
\usetikzlibrary{arrows,shadows}
\usepackage{pgf-umlsd}

\usepackage{tkz-euclide} 
\usepackage{pgf-pie}

\newcommand{\tikzcircle}[2][red,fill=red]{\tikz[baseline=-0.5ex]\draw[#1,radius=#2] (0,0) circle ;}%

\begin{document}

\title{Exploring Web3 From the View of Blockchain}

\subtitle{(Tech Report)}
\renewcommand\rightmark{}
\renewcommand\leftmark{Web3 in Blockchain}
\newcommand{\romannum}[1]{\romannumeral #1}


\author{Qin Wang$^{\star}$\inst{4}, Rujia Li$\thanks{These authors have contributed equally to the work. \\ Affiliations are ordered by the text length. }$\inst{2,3}, Qi Wang\inst{2}, Shiping Chen\inst{4}, \\ Mark Ryan\inst{3}, Thomas Hardjono\inst{1} }
\authorrunning{Qin Wang, Rujia Li, et al.}
\titlerunning{Tech Report on Web3}

\institute{
\text{MIT Connection Science, Massachusetts Institute of Technology,} USA
\and
\text{Southern University of Science and Technology}, China
\and
\text{University of Birmingham}, United Kingdom
\\
\and
\text{CSIRO Data61}, Australia
}

\maketitle           

\begin{abstract}
Web3 is the most hyped concept from 2020 to date, greatly motivating the prosperity of the Internet of Value and Metaverse. However, no solid evidence stipulates the exact definition, criterion, or standard in the sense of such a buzzword. To fill the gap, we aim to clarify the term in this work. We narrow down the connotation of Web3 by separating it from high-level controversy argues and, instead, focusing on its protocol, architecture, and evaluation from the perspective of blockchain fields. Specifically, we have identified all potential architectural design types and evaluated each of them by employing the scenario-based architecture evaluation method. The evaluation shows that existing applications are neither secure nor adoptable as claimed. Meanwhile, we also discuss opportunities and challenges surrounding the Web3 space and answer several prevailing questions from communities. A primary result is that Web3 still relies on traditional internet infrastructure, not as independent as advocated. This report, as of June 2022, provides the first strict research on Web3 in the view of blockchain. We hope that this work would provide a guide for the development of future Web3 services.

\keywords{Blockchain \and Web3 \and Internet of Value \and Architecture}
\end{abstract}

\section{Introduction}
Web3, also known as Web 3.0 or decentralized web, hits the cryptocurrency markets and blockchain communities from 2020 to date \cite{web3soul}\cite{csnreport}. It has become the most prevailing term in the recent period of blockchain prosperity. The concept, proposed by Wood \cite{woodweb3}, promises to provide distributed internet services without trusted third parties (TTP), thereby offering users more control over their data. The very primary principle shared by Web3 applications is that users can hold the data with full control, covering identifiers/tokens/ownership/etc., rather than being managed by centralized organizations as in Web1 and Web2 (cf. Appendix A). With the emphasis on \textit{decentralization}, Web3 moves data away from these central authorities and establishes applications and services surrounding blockchain technologies. Ideally, Web3 developers do not need to build applications on top of a single server (processing business logic) or database (storing user data). Instead,  Web3 applications are deployed on decentralized networks such as blockchain platforms or related distributed systems hosted by many peer-to-peer (P2P) servers.

The conversion from central authorities to the blockchain, more than handing over the ownership back to users, brings many non-functional benefits of being (i) \textit{open}: Web3 data is stored in an open network developed by public communities. Also, Web3 applications are executed in a global view towards the public, making data permanently visible to all the participants; (ii) \textit{trustless}: a user can build connections or exchange assets to an unfamiliar user without the reliance of a trusted third party; (iii) \textit{permissionless}: users' identities are no longer tied to any specific platform and users' activities are free, which do not need an authorization from a governing entity; (iv) \textit{anonymous}: users can obtain partial anonymity through using multiple pseudonyms or off-chain storage; and of (v) \textit{high availability}: Web3 provides a high availability architecture, which reduces the probability of the server crash or single point failure;  (vi) \textit{compatibility}: Deployed services and applications are not limited in the Ethereum ecosystem, but applicable to all competitive blockchains \cite{dappradar,mathdapp} like Avalanche \cite{avax}, Solana \cite{solana}, Binance Smart Chain \cite{bsc} or Oasis Network \cite{rose}, indicating that Web3 can integrate both existing public-chain ecosystems and incoming systems.

Web3 also improves user experience in using web stack technologies and thus promotes the development of blockchain technologies. Web3 helps to build a complete user experience that can better serve newcomers. Users, from the current view, only need to \textbf{\textit{connect the wallet}} to their targeted sites. These sites contain the activities that a user would love to spend time on, which can be specified in different types, such as asset swaps (in DEXes \cite{xu2021sok}), games \cite{axieinfinity}, and trades (of NFTs \cite{wang2021non}). Meanwhile, users can obtain a portion of revenues by keeping active in these projects. The contributions are measured in a wide range covering spent time, interactions, deposits, staking, etc. Users can receive fair rewards as well as add more liquidity to the Web3 ecosystem. In this sense, Web3 has greatly extended the boundary of blockchain, incentivizing more users to participate in the games. The evidence is, as in Dec 2021, the total locked value (TLV) of Web3-related smart contact contract has reached a peak with 193.14 billion USD (more details in Sec.\ref{subsec-impact}). Such a high capitalization market further stimulates the prosperity of techniques surrounding blockchain (cf. Tab.\ref{tab-projects}). 

However, despite the fact that Web3 has drawn much attention, clear statements of \textit{what is Web3} and \textit{how Web3 is designed} are even absent. Several studies provided their investigation on the consensus level~\cite{liu2021make} but failed to cover a full view of other components and architectural designs that are equally important to Web3. Unclear definitions and non-agreed consensus indicate that Web3 either is just a hyped concept without practical development or, has more than one single direction for the development. In this technical report, we avoid such high-level discussions, narrowing down the scope to the Web3 architecture and its relation with blockchain. We dig into its building process, investigate its usage in existing projects, and deconstruct the protocol into separated components. With this in the arm, we analyze and evaluate \textit{state of the art} Web3 solutions from perspectives of their design patterns and properties. We further extend our scope to the entire Web3 space (still rooted in blockchains) by discussing their impacts, opportunities, and challenges. To our knowledge, this work provides the first in-time study on Web3, reviewing and exploring the wild Web3 solutions.

\begin{itemize}
\item[-] We investigate existing Web3 projects and extract a succinct backbone model with participating roles and operating workflow to demonstrate the major processes. This provides a loose general model for easy adoption, as well as highlights the core steps to establish a Web3 service.

\item[-] We classify a total of twelve architectural designs of Web3 to reflect the operating mechanism of typical Web3-based applications. We decouple a full Web3 service into three-layered components according to the data workflow. Data in each component can be either operated by the ways of \textit{on-chain}, \textit{off-chain} or \textit{hybrid}. The identified design types can well present all potential combinations, covering a wide range of Web3 applications and services.

\item[-] We accordingly evaluate each architecture based on different property metrics that are grabbed from classic blockchain systems. The evaluation analyses the architectures from multi-dimensional properties (cf. Tab.\ref{tab-archievalua}) when adapting to real applications. We also discuss which participating entity can gain the most benefits under different types. The evaluations directly help software architects to evaluate and compare Web3 solutions and further design their architectural frameworks.

\item[-] We extend our scope from architectural design to the entire Web3 space by discussing its impacts, opportunities and challenges to predict the potential future directions. We show several promising fields that may be inspired by Web3-related technologies, and also point out the open challenges needed to be addressed in the long term. Meanwhile, based on our analyses and evaluations, we answer several questions that are frequently asked by the communities as our closing results to conclude this work.

\item[-] Further, we provide much more surrounding knowledge that is related to Web3, covering its basic primitives, referenced token standards, and adopted programming languages. Necessary primitives cover both architectural design in conventional web services and layered components in classic blockchain systems. The knowledge lays the foundation of Web3, presenting the common components in different blockchain-based Web3 projects. 

\end{itemize}

The rest of this paper is organized as follows. Section \ref{sec-protocol} presents a typical Web3-based protocol, together with the inducted security model. Section \ref{sec-archi} presents our architecture extraction and design, as well as gives the criterion used to capture the features of Web3 related applications. Section \ref{sec-evaluation} details our evaluation towards architectural designs, providing practical adaption under different scenarios. Section \ref{sec-discussion} discuss current impacts, opportunities and challenges. Finally, Section \ref{sec-closing} gives our closing results and Section \ref{sec-conclu} concludes this work. Notably, we provide more knowledge of Web3: Appendix A gives fundamental primitives surrounding Web3 and an abstract Web3 workflow. Appendix B and C summarize existing token standards and EVM-operated programming languages, respectively. Appendix D shows a small corpus of \textit{in the wild} Web3 projects.

\section{Web3 Protocol}
\label{sec-protocol}

This section extracts the protocol, security model, and a running example of Web3 systems between user clients and the blockchain infrastructure (together with the protocol workflow, referring to Fig.\ref{fig-workflow} in Appendix A.3). Before our investigation, we clarify the entities that participate in Web3 protocols.

\smallskip
\textbf{Involved Entities.} The Web3 protocol mainly consists of three types of roles, namely, the \textit{Web3 user}, \textit{service provider}, and the \textit{blockchain maintainer}. The Web3 user is a data owner who can propose the request by sending a transaction from the client. The service provider offers the user an on-chain interface and the service that can both process the request and interact with blockchain platforms. The blockchain maintainer provides an operating environment for smart contracts that take the task of processing business and data storage.

\smallskip
\textbf{{General Construction.}}
To launch the Web3 service, a Web3 user needs to first establish his distributed identity by creating an address or account on-chain at the client. Then, the user login the applications using his address-based identity. Subsequently, the application will interact with the blockchain for data computation and data storage. Here, we present the detailed protocol.

\noindent\hangindent 1em \textbf{\textit{\,\, Identity Creation} ($\mathsf{\Pi_1}$).} This algorithm takes as input the security parameter $\kappa$, and outputs a blockchain address/account $addr$. The address provides a public and open-source index that can be associated with an existing identifier. Also, this address is used as an identifier for sending or receiving the cryptocurrency. It covers three sub algorithms, \textit{private key generation}, \textit{public key generation} and \textit{address generation}.

\begin{equation*}
\centering
\begin{aligned}
sk \gets \mathsf{KeyGen_{sk}}(1^{\kappa}), \\
pk \gets \mathsf{KeyGen_{pk}}(sk),\\
addr \gets \mathsf{AddressGen}(pk).
\end{aligned}
\end{equation*}

\noindent\hangindent 1em \quad The algorithm is run by a local Web3 client (e.g., client-based wallet, browser extension-based wallet). Note that the sub-algorithms $\mathsf{KeyGen_{sk}}$, $\mathsf{KeyGen_{pk}}$ and, $\mathsf{AddressGen}$ have different manifestations. For example, for $\mathsf{AddressGen}$, Bitcoin addresses use the base-58 encoding\footnote{https://tools.ietf.org/id/draft-msporny-base58-01.html}, while Ethereum addresses adopt the base-16 encoding algorithms.

\noindent\hangindent 1em \textbf{\textit{\,\, Transaction Generation} ($\mathsf{\Pi_2}$).} This algorithm takes as input the user's private key $sk$, the transaction $metadata$, the transaction $payload$, and accordingly outputs an enveloped transaction $Tx$.

\begin{equation*}
\begin{aligned}
sig \gets \mathsf{Sign}(sk, metadata, payload), \\
Tx \gets \mathsf{TranGen}(sig, metadata, payload).
\end{aligned}
\end{equation*}

\noindent\hangindent 1em \quad This algorithm is run by a Web3 client that receives the request from users. $sk$ is used to generate a signature. $metadata$ refers to transaction-related data, such as the transaction receiver and transaction nonce. $payload$ points to the contract method to be involved and the data to be stored. 

\noindent\hangindent 1em \textbf{\textit{\,\, Contract Execution} ($\mathsf{\Pi_3}$).} This algorithm takes as input the transaction $Tx$, and current state $\overline{s}$, and smart contract $contract$, and accordingly outputs the transferred state $s$.
\begin{equation*}
s \gets \mathsf{ContractExec}(\overline{s}, Tx, contract).
\end{equation*}
\noindent\hangindent 1em \quad This algorithm is run by the blockchain maintainers in the network. The transited state $s$ includes a reward or payment used to incentive the participating nodes. The contract $contract$ contains the logic of upper-layer applications, which defines how the state changes happening on the blockchain.

\noindent\hangindent 1em \textbf{\textit{\,\, State Consensus} ($\mathsf{\Pi_4}$).} This algorithm takes as input the transaction $Tx$, the smart contract $contract$ and the current state $s$ to be transited, and outputs the confirmed state $s^{\prime}$, and the confirmed transaction $Tx^{\prime}$.

\begin{equation*}
(s^{\prime}, Tx^{\prime}) \gets \mathsf{Consensus}(s, Tx, contract).
\end{equation*}

\noindent\hangindent 1em \quad  This algorithm is run by blockchain maintainers. The term \textit{confirmed} indicates that the proposed block (containing $s$ and $Tx$) has been agreed upon by sufficient maintainers in the network. The  threshold of confirmation depends on specific algorithms (e.g., 51\% in PoW \cite{nakamoto2008bitcoin}, 2/3 in BFT algorithms \cite{castro1999practical}).

\noindent\hangindent 1em \textbf{\textit{\,\, State Retrieval} ($\mathsf{\Pi_5}$).} This algorithm takes as input the user's address $addr$ and outputs the transaction $Tx^{\prime\prime}$ and the related state $s^{\prime\prime}$.

\begin{equation*}
(s^{\prime\prime}, Tx^{\prime\prime}) \gets \mathsf{Retrieval}(addr,  contract).
\end{equation*}

\noindent\hangindent 1em \quad  This algorithm is run by a Web3 client who aims to retrieve the state.

\smallskip
\noindent\textbf{Secure Web3 Protocols}. Based on the extracted general construction, a secure Web3 protocol is informally defined as follows.

\smallskip
\begin{defi}[Secure Web3 Protocol]
A Web3 scheme $\mathsf{S}$ is secure, if for all blockchain address $addr$ and the initial state $\overline{s}$ outputted from Phase $\mathsf{\Pi_1}$ to Phase $\mathsf{\Pi_5}$, it holds that,

\begin{equation*}
\left[
\begin{array}{ll}
\begin{aligned}
     sig \gets \mathsf{Sign}(sk, metadata, payload) \\
Tx \gets \mathsf{TranGen}(sig, metadata, payload) \\
s \gets \mathsf{ContractExec}(\overline{s}, Tx, contract) \\
 (s^{\prime}, Tx^{\prime}) \gets \mathsf{Consensus}(s, Tx, contract) \\
\end{aligned}
\end{array}
\right]  \Rightarrow (s^{\prime}, Tx^{\prime})
\end{equation*}

\begin{equation*}
\left[
\begin{array}{ll}
\begin{aligned}
(s^{\prime\prime}, Tx^{\prime\prime})\gets \mathsf{Retrieval}(addr,  contract) \\
\end{aligned}
\end{array}
\right]  \Rightarrow (s^{\prime\prime}, Tx^{\prime\prime})
\end{equation*}


where the output satisfies:
\begin{align*}
(s^{\prime}, Tx^{\prime}) = (s^{\prime\prime}, Tx^{\prime\prime})
\end{align*}

\end{defi}

The definition states that a Web3 protocol is deemed to be secure if a user can successfully retrieve the correct state and transaction on the blockchain at any time after the block confirmation. Our security model is based on the assumption of a \textit{robust} blockchain \cite{garay2015bitcoin} and the corresponding security is guaranteed by \textit{persistence} and \textit{liveness}~\cite{garay2017bitcoin}. The \textit{persistence} means whether different nodes have the same view at a specific height of blocks, while the \textit{liveness} focuses on whether a block can be eventually buried deep enough (without the possibility of being reversed) in the valid longest chain. We omit other types of chain structure, such as the directed acyclic graph (DAG) \cite{wang2020sok}.

\begin{figure}[!hbt]  
    \centering
    \includegraphics[width=0.9\linewidth]{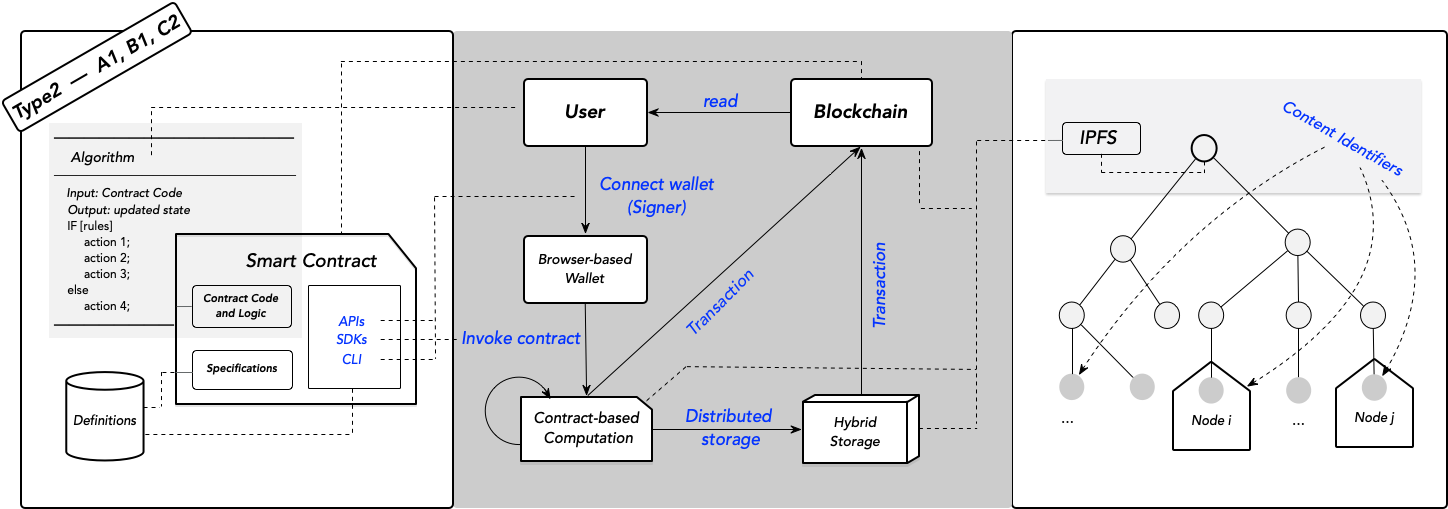}
   \caption{An Instance of Web3 User Case}
   \label{fig-instance}
\end{figure}

\textbf{A Running Example.} 
We provide an instance to describe how a Web3 system works. The case is straightforward: assume that a user wants to sell his self-created NFT painting on-chain (cf. Fig.\ref{fig-instance}). Firstly, he needs to connect his wallet to the related website and send the request to mint an NFT on his targeted blockchain platforms \cite{wang2021non}. The action of \textit{mint} requires invoking APIs that enable interactions with smart contracts. The contract is executed according to its predefined specifications and token standards (ERC721, see details in Appendix B). After processing the logic, the raw picture is stored in an external decentralized storage network, e.g., IPFS \cite{ipfs}, due to its high-resolution format. IPFS separates the raw data into pieces and distributes them into different nodes, labelling each of them with a content identifier (CID). The returned identifiers are recorded on-chain for further queries from users. Once completing the entire logic, the seller can trace the transaction and his NFT on the blockchain. The buyer with willingness can buy the seller's NFT on Opensea \cite{opensea}, and pay the prices as stated. The ownership of this NFT will be automatically transferred to the buyer once completing the payments.

\section{Architectural Design} 
\label{sec-archi}

The architecture is the foundation of software systems. It defines the system structure and external-world interface via underlying components and their relationship. This section provides our architectural design for Web3 systems. 

\subsection{Design Principle}

\begin{figure}[h!]
    \centering
    \includegraphics[width=1\linewidth]{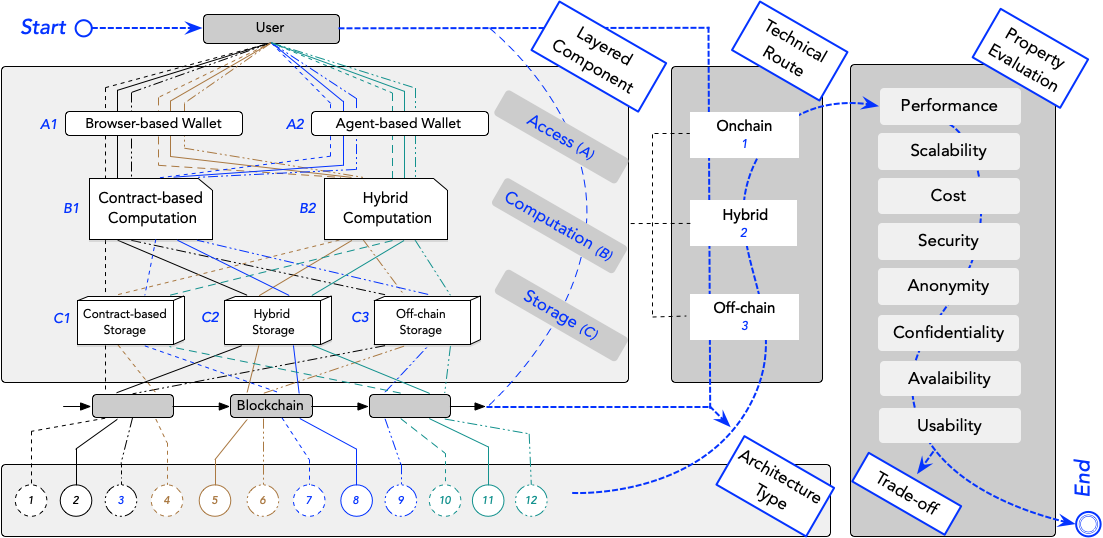}
   \caption{\textbf{\textit{Framework.}} We have first identified three major components according to the data workflow: access, computation, and storage (cf. \textit{left Square}). Then, we extract three technical routes that reflect the ways of data processing in blockchain systems: on-chain, off-chain, and hybrid (\textit{medium Square}). Based on the two principal dimensions, we summarise a total of $12$ types of potential Web3 architectural designs (\textit{bottom Square}). With this in the arm, we analyze every type in terms of different property metrics (\textit{right Square}) and point out design trade-offs. Notably, we also provide a guide map by using \textit{blue dotted line}. }
   \label{fig-model}
\end{figure}

We stay aligned with the mainstream recognition of Web3 and accordingly structure our report by progressively presenting the ways to build a Web3 service and the featured properties in different designs. Based on plenty of investigation towards \textit{in the wild} Web3 projects, we classify them into a compact framework (cf. Fig.\ref{fig-model}), which thus allows the audience to explore the conceptual design space and evaluate different design options. 

Specifically, we delineate two basic axes, namely, \textit{decoupled component} and \textit{technical route}, to explore the subtle differences in each cross from orthogonal directions. For the former, we abstract three critical components, including  \textit{client}, \textit{compuation} and \textit{storage}, as our characterisation. Our decoupling is based on the data workflow, which contains three corresponding phases (A-C in Fig.\ref{fig-model}), \textit{data access}, \textit{data computation} and  \textit{data storage}, to describe how data transfers under a normal operating logic. For the latter, we extract three technical routes of implementing a blockchain system: \textit{on-chain} processing, \textit{off-chain} processing, and their \textit{combinations} (hybrid).

By interweaving two axes, we can find that each component can be categorized into sub-items according to its on-chain/off-chain processing route. Based on that, we identify a total of twelve types of architectural designs to show different purposes and usages among a large bunch of projects. Moreover, we capture several essential properties that are frequently mentioned in classic blockchain systems to evaluate the proposed design types. This framework captures major architectural characteristics and related properties of each type, helping Web3 users and software architects to choose the proper design for their products. We give the detailed definitions of each item as follows. 

\subsection{Decoupled Components}
\label{subsec-compo}

We first investigate how data flows from a user to its backend server. In such a procedure, we have totally extracted three major processes to normal system operations: \textit{data access}, \textit{data computation} and \textit{data storage}. Such procedures describe a typical workflow of blockchain systems.

\begin{itemize}
 \item[-] \textbf{\textit{Data Access.}} To start a Web3 service, a user first needs to send his request through a client. Typically, the request is formatted as a transaction, and accordingly, the client is instantiated as a wallet\footnote{We omit the sub-categories of different wallet types, such as whether utilizing SPV or hardware. More details refer to \cite{chatzigiannis2021sok}\cite{karantias2020sok}.}. Users enter the network by connecting their wallets to a specific website.
 
  \item[-] \textbf{\textit{Data Computation.}} After a successful connection, the requests from users are parsed into different pieces of logic, being transmitted into backend blockchain platforms. Blockchain operates the logic and decoded methods via smart contracts according to their predefined specifications. 
 
 \item[-] \textbf{\textit{Data Storage.}} The executed data, at last, should be permanent storage. In normal cases, small size data can be directly recorded inside transactions, whereas in some cases, large-scale data requires external storage. The key to maintaining the data integrity in off-chain storage is to add a \textit{hook} that connects the on-chain (such as a hash) and off-chain data.
 
\end{itemize}

Then, based on descriptions of data flow, we abstract three fundamental components in a full-functional Web3 service by decoupling existing solutions. Merging with our different data processing routes, we categorize the decoupled component with \textit{on-chain}, \textit{off-chain} and \textit{hybrid} options (selected by two or all of them). We give their definitions as follows (and more details in Sec.\ref{subsec-component}). To be noted, as Web3 is featured by decentralization, at least one of the components should be operated on-chain, where no gatekeeper can fully control the network. 

\smallskip  
\textbf{Client.} Clients are used to receiving requests from users. Classic blockchain systems, on a small scale of requests, can use a single browser as the client to connect with the wallet. However, if the request sharply increases within a short time frame, an agent is required to process an instantaneous flow of requests.

\begin{itemize}
 \item[-] \textbf{\textit{Browser-based Wallet.}} A browser-based wallet is an intuitive solution for users to adopt the Web3 service. Users only need to install an extension tool in the browser and import their private key to this embedded wallet. When browsing a Web3-supported website, users can directly put the button of \textit{connect the wallet}, and all the clicked functions in this website will invoke the backend methods through APIs under the user's account.
 
 \item[-] \textbf{\textit{Agent-based Wallet.}} An agent solution is to enable batch processing when confronting a high-density situation of requests from users. Similar to traditional Web1/Web2, users should first grant a trusted agent with proper permission. The authentication procedure is executed once users have formally started to register with the agents.
\end{itemize}

\smallskip
\textbf{Computation.} Operating a Web3 service built on smart contracts requires great computing power as the running environment. Purely on-chain computation is instant but costly. A hybrid computation that moves partial computational tasks off-chain is another solution.

\begin{itemize}
 \item[-] \textbf{\textit{Contract-based Computation.}} All the computations are performed on-chain through smart contracts. Here, variables in smart contracts are allowed to be revoked and modified based on the predefined contract instructions.
 
 \item[-] \textbf{\textit{Hybrid Computation.}} A partial proportion of computations are executed on-chain, while the rest of the computations off-chain. Generally, verification mechanisms are required in this method in order to guarantee integrity between on-chain and off-chain data.
 
\end{itemize}
\smallskip
\textbf{Storage.} Similar to any services applied to Web1/Web2,  applications and services in Web3 also require enough space for storage. Data in small size, such as ordinary transaction payload, can be stored directly on-chain. In contrast, for the large size, or complicatedly formatted data, like streaming video or raw audio, it should be (partially) stored to external deceives or providers. In this sense, we summarize three types of data storage. 

\begin{itemize}
 \item[-] \textbf{\textit{On-chain Storage.}} All the data is stored on-chain, as a state of the smart contract. Any changes will be publicly recorded and can be verified by the entire network. Theoretically, this is the best way to prevent malicious nodes from destroying data due to its total transparency and accountability.
 
 \item[-] \textbf{\textit{Off-chain Storage.}} It refers to storing data in off-chain networks where only the hash roots are recorded online (e.g., IPFS \cite{ipfs} or Swarm \cite{swarm}). The solution requires an additional process for verification, such as proof-of-existence, to prove data integrity.
 
 \item[-] \textbf{\textit{Hybrid Storage.}} This is a mixed way that combines both on-chain and off-chain storage, which means storing raw data off-chain (like video and audio) and light metadata (such as account history and certificates) on-chain. The proportion of these solutions is adjusted according to specific scenarios. 
\end{itemize}

\subsection{Architectural Designs}\label{subsec-design}

We have identified totally twelve types of architectural designs towards Web3 services (cf. Architectural Type $1-12$ in Fig.\ref{fig-model}). Specifically, we denote each design with \textit{Type}, and it is made up by a three-element tuple $(A_a,B_b,C_c)$ where $a,b \in \{1,2\}$, and $c \in \{1,2,3\}$. Here, the elements $A,B,C$ separately specify the components of \textit{client}, \textit{computation} and \textit{storage}, while the subscripts $1,2,3$, respectively, represent \textit{on-chain}, \textit{hybrid} and \textit{off-chain} as defined before. Based on different combinations, we provide the potential architectural designs as follows.

\begin{itemize}
    \item[-] \textit{Type1} $(A_1,B_1,C_1)$, \textit{Type2} $(A_1,B_1,C_2)$, \textit{Type3} $(A_1,B_1,C_3)$, which are shown in the \textit{black line} at the left square in Fig.\ref{fig-model};
    
    \item[-] \textit{Type4} $(A_1,B_2,C_1)$, \textit{Type5} $(A_1,B_2,C_2)$, \textit{Type6} $(A_1,B_2,C_3)$ (\textit{\textcolor{brown}{brown line}});
    
    \item[-] \textit{Type7} $(A_2,B_1,C_1)$, \textit{Type8} $(A_2,B_1,C_2)$, \textit{Type9} $(A_2,B_1,C_3)$ (\textit{\textcolor{blue}{blue line}});
     
    \item[-] \textit{Type10} $(A_2,B_2,C_1)$, \textit{Type11} $(A_2,B_2,C_2)$, \textit{Type12} $(A_2,B_2,C_3)$ (\textit{\textcolor{teal}{teal line}}).
\end{itemize}

In each design type, smart contracts play the most essential role in entire services that are responsible for on-chain calculations. \textit{Type1} is the most simplified architectural design amongst those combinations. A typical instance is known as today's Ethereum platform. To fit for more complicated cases, external resources should be equipped to the system design, such as the agent in \textit{Type7-12} or the off-chain storage in  \textit{Type2/3}. Notably, the running NFT example in Sec.\ref{sec-protocol} is a based on the \textit{Type2} design.

\section{Architecture Evaluation}
\label{sec-evaluation}
In this section, we present the evaluation criteria and evaluate each type of architecture. Our evaluation covers the properties that are summarized from classic blockchain systems and benefits to different participating parties. 

\subsection{Evaluation Criteria}
\label{subsec-criteria}

In the past decades, many evaluation frameworks have been proposed. Among them, the \textit{architecture trade-off analysis method} (ATAM) \cite{kazman1998architecture} is a well-known one. We partially adopt ATAM to assess the Web3 architecture, integrating with the blockchain properties and beneficial parties. In particular, we first select a concrete scenario, refine quality attributes into this scenario, and then present potential concerns related to the quality requirement.

\smallskip
\textbf{ATAM Method.} The architecture trade-off analysis method is initially used in software development to recommend the most suitable architecture for a specific system. The method majorly discusses the trade-offs between different design types and their sensitivity points for risk mitigation. A classic ATAM process consists of three critical aspects: \textit{extract quality attributes}, \textit{identify beneficial parties,} and \textit{analyse in specific scenarios}. Quality attributes are non-functional requirements to specify how well software should be done. In the context of Web3, we adopt the properties extracted from classic blockchain systems as quality attributes. Then, we analyse the system entities regarding their potential benefits and losses. Meanwhile, we still use a similar NFT instance (compared to the running example in Sec.\ref{sec-protocol}) as our analysed scenario.  

\smallskip
\textbf{Property Metrics.}\label{subsec-property}
We capture several major properties that are used to evaluate classic blockchain systems. These metrics depict the architecture from multiple dimensions, and we summarize them as follows. 

\begin{itemize}
    \item[-] \textit{Performance.} Performance is used to show the maximum rate of confirmed transactions during a specified duration. In the context of blockchain, performance is measured by transaction per second (TPS), which is a widely adopted way to describe the throughput of a system.
    
    \item[-] \textit{Scalability.} Scalability refers to the ability to process transactions along with the increased scale of networks, which is reflected by the increase or decrease in its performance (measured in transaction per second--TPS).
    
    \item[-] \textit{Cost.} Each transaction in VM-engined blockchain platforms requires certain transaction fees. This is necessary because a total zero friction (free of fees) will result in a loop logic error \cite{kalra2018zeus} in smart contract operations.
    
    \item[-] \textit{Security.} Security is a broad term. We limit the scope of security to data-level security. The data should be protected for its integrity and prevented from unauthorized access or faking, or compromising.
    
    \item[-] \textit{Anonymity.} Anonymity refers to hiding the identities of acting persons. In the context of blockchain, anonymity mainly indicates utilizing cryptographic schemes to break the linkage between physical entities with virtual public addresses~\cite{Li2022SoKTC}. An adversary cannot learn useful information about any specific parties related to the on-chain assets.
     
    \item[-] \textit{Confidentiality.} Confidentiality refers to hiding the sensitive data that is presented on transactions, including their asset values, useful information in payloads, or contract states. Confidentiality focuses on states rather than entities. Notably, privacy covers both anonymity and confidentiality, where an adversary cannot learn any useful knowledge from public transactions.
    
    \item[-] \textit{Availability.} Availability is used to measure the probability of a system running abnormally with the existence of potential failure. It indicates the stability of a blockchain system.
    
    \item[-] \textit{Usability.} Usability in the context of Web3 follows the same connotation as the general computer system, which measures the satisfaction of a specific user in a specific context when using a specific product.
    
\end{itemize}

\smallskip 
\textbf{Stakeholders.}
A major task in ATAM evaluation is to analyse a system's capability for satisfying the stakeholder requirements~\cite{bass2003software}. The stakeholders in the context of our Web3-based evaluation indicate the potential beneficial parties in the system, being aligned with \textit{involved entities} stated in Sec.\ref{sec-protocol}, rather than the person holding assets in proof of stake (PoS) protocols \cite{kiayias2017ouroboros}. Three main stakeholders are identified as our evaluation target.

\begin{itemize}

\item[-] \textit{Web3 user.} A Web3 user is an entity that owns the data. Also, the user is an entity to initiate transactions. The user should accordingly obtain a fair proportion of rewards in the forms such as tokens, badger, etc.


\item[-] \textit{Service provider.} An application owner uses blockchain to provide publicly accessible services for users. The service provider may obtain revenues from both sides: user payments and blockchain platform rewards.

\item[-] \textit{Blockchain maintainer.} Blockchain maintainer provides a running environment for executing the smart contract and storing the blockchain data. The main reward for maintainers comes from the cryptocurrency rewards.

\end{itemize}

\subsection{Architecture Evaluation}
\label{subsec-evaluation}
Then, we start to evaluate each architectural design. We give a specific scenario to narrow down the scope of the high-level description and analyze each architecture regarding its properties and stakeholders.

\textbf{Scenario.} Non-Fungible Token \cite{wang2021non} is a blockchain-based cryptocurrency that allows proving the ownership. In our scenario, an NFT designer, Alice, wants to sell her NFT to Bob by paying cryptocurrencies. Even though this application focuses on a specific domain, we argue that the core requirements are likely to stand across different Web3 applications. We apply different architectures to this scenario, discussing both advantages/disadvantages and trade-offs, to help interested readers consider their final decisions on Web3 architectures.

\smallskip
\textbf{\textit{A1: An Ideal Web3 Architecture.}} An ideal Web3 infrastructure should eliminate all the middle man, without centralized databases or computation devices. The blockchain provides a decentralized platform for data computation and storage while bringing a native token-based economy. Under this architecture, Alice uploads the raw NFT data, the price information, and NFT auxiliary information to a blockchain-based smart contract through a wallet. Then, the blockchain nodes reach an agreement on the data. Next, Bob sends a transaction to invoke the contract for buying NFT. Finally, the NFT ownership transfers to Bob, and Alice obtains the payments. The architecture puts all the computation and storage on the blockchain, bringing security and availability for Web3 users. The blockchain operates on a collection of decentralized nodes, which guarantees the availability of on-chain data. Meanwhile, once blockchain nodes reach an agreement for the received transactions, data becomes immutable. The related architectural design in Sec.\ref{sec-archi} is \textit{Type1}. 

However, this architecture confronts some drawbacks, which mainly center around being impractical when adapting to today’s Web3 infrastructure.  Firstly, uploading and replicating the data to all blockchain nodes in a peer-to-peer network is time-consuming. Website interactions are slower where the back-ended state transitions need confirmation and propagation throughout the network. Meanwhile, scalability is a long-term issue existing in blockchain systems due to intrinsic decentralization. The architecture is closely integrated with underpinned blockchain systems, which, as a result, will face a similar issue. Moreover, the computation and storage in the blockchain are expensive regarding monetary costs. Worse still, the fees increase with the growing size of payload data.

Beyond that, this architecture raises problems about data privacy. On the one hand, the code and state of smart contracts are transparent by default. Any changes in the contracts are immediately visible, not just to the blockchain nodes but also to anyone outside blockchains. The over-publicity issue hinders the development of Web3 applications, making them hard to be adopted for privacy-critical applications. On the other hand, users' identities lack anonymity. Without an explicit approach taken to protect the user's addresses, virtual addresses on blockchains and physical identities are linkable with the help of analytical tools and big data, even if users can use multiple pseudonyms.

\smallskip
\textbf{\textit{A2: Agent-login Architecture.}} The ideal Web3 solutions, singly relying on the original blockchain technology, confront the issues of poor scalability and high cost. These shortcomings force the user to adopt the alternatively centralized computation or storage outside the blockchain. We gradually proceed from layer to layer. Firstly, we address the congestion issue at the \textit{access} layer. The agent-login design provides a scalable approach to handling many transactions by merging multiple executions. A group of users can access the blockchain platform with one agent account. In particular, in our scenario, multiple NFT-selling and buying transactions are bundled into one transaction. An agent anchors this transaction to the blockchain later for the final agreement. Centralized exchanges, such as Coinbase~\cite{coinbase} or Binance~\cite{binance}, act as agents similar to this design. Many Web3 participants rely on custodial wallets for easy access and management. This architecture can concurrently process transactions without waiting for the blockchain's confirmation, which brings high performance, good scalability, and low cost. The related architectural designs include \textcolor{blue}{\textit{Type7-9}}.
The main drawback of this solution is centralization, which may cause unnecessary issues. For example, a malicious agent may hide the user's transactions without sending them to the blockchain. Worse still, such action is hard to trace due to the lack of transparent evidence.

\begin{table}[!hbt]
    \centering
\caption{Web Architecture Evaluation}
\label{tab-archievalua}
\resizebox{\linewidth}{!}{ 
\begin{tabular}{lccccccccc|cccc@{}}
\toprule
\multicolumn{2}{c}{} & \multicolumn{7}{c@{}}{\textbf{Property}} & \multicolumn{5}{c@{}}{\textbf{Stakeholders}}\\
\cmidrule(l){3-14}
\multicolumn{2}{c}{\textbf{Architecture}}  & \rotatebox[origin=c]{65}{Performance} & \rotatebox[origin=c]{65}{Scalability} & \rotatebox[origin=c]{65}{Gas Cost} & \rotatebox[origin=c]{65}{Security}  &
\rotatebox[origin=c]{65}{Anonymity} &
\rotatebox[origin=c]{65}{Confidentiality} &
\rotatebox[origin=c]{65}{Availability} &
\rotatebox[origin=c]{65}{Usability} &
\rotatebox[origin=c]{65}{} &
\rotatebox[origin=c]{65}{Web3 User} &
\rotatebox[origin=c]{65}{Service Provider} &
\rotatebox[origin=c]{65}{BC Maintainer} \\ 
\cmidrule(l){1-14}

& \multicolumn{1}{l}{ $A_1$,$B_1$,$C_1$--\textit{Type1} } & 
\tikzcircle[black, fill=white]{2pt} 
& 
\tikzcircle[black, fill=white]{2pt} 
&
\tikzcircle[black, fill=white]{2pt}  & \tikzcircle[black, fill=white]{2pt}  & \tikzcircle[black, fill=white]{2pt}  &  \tikzcircle[black, fill=white]{2pt} 
& \tikzcircle[black, fill=white]{2pt} 
& \tikzcircle[black, fill=white]{2pt}  & & \tikzcircle[black, fill=white]{2pt}  & \tikzcircle[black, fill=white]{2pt}   & \tikzcircle[black, fill=white]{2pt}   \\ 

&  \multicolumn{1}{l}{ $A_1,B_1,C_{2/3}$--\textit{Type2/3} } &
\tikzcircle[green, fill=green]{2pt}
\tikzcircle[green, fill=green]{2pt}
&  
\tikzcircle[green, fill=green]{2pt}
\tikzcircle[green, fill=green]{2pt}
&  
\tikzcircle[green, fill=green]{2pt}
& 
\tikzcircle[red, fill=red]{2pt} 
& 
 \tikzcircle[black, fill=white]{2pt} 
&
\tikzcircle[green, fill=green]{2pt}
\tikzcircle[green, fill=green]{2pt}
&
\tikzcircle[red, fill=red]{2pt}
\tikzcircle[red, fill=red]{2pt}
& 
\tikzcircle[green, fill=green]{2pt}
& & + & - & - \\

\cmidrule{2-2}
& \multicolumn{1}{l}{$A_1$,$B_2$,$C_1$--\textit{Type4}} & 
\tikzcircle[green, fill=green]{2pt}
& 
\tikzcircle[green, fill=green]{2pt} 
& 
\tikzcircle[green, fill=green]{2pt} 
& 
\tikzcircle[red, fill=red]{2pt} 
\tikzcircle[red, fill=red]{2pt} 

& 
 \tikzcircle[black, fill=white]{2pt} 
& 
\tikzcircle[green, fill=green]{2pt}
\tikzcircle[green, fill=green]{2pt}
&
\tikzcircle[red, fill=red]{2pt}
\tikzcircle[red, fill=red]{2pt}
& 
\tikzcircle[green, fill=green]{2pt}
& & + & -& - \\

& \multicolumn{1}{l}{$A_1$,$B_2$,$C_{2/3}$--\textit{Type5/6}} & 
\tikzcircle[green, fill=green]{2pt} 
\tikzcircle[green, fill=green]{2pt}
& 
\tikzcircle[green, fill=green]{2pt} 
\tikzcircle[green, fill=green]{2pt}
& 
\tikzcircle[green, fill=green]{2pt} 
\tikzcircle[green, fill=green]{2pt} 
& 
\tikzcircle[red, fill=red]{2pt} 
\tikzcircle[red, fill=red]{2pt} 
\tikzcircle[red, fill=red]{2pt} 

& 
 \tikzcircle[black, fill=white]{2pt} 
& 
\tikzcircle[green, fill=green]{2pt}
\tikzcircle[green, fill=green]{2pt}
&
\tikzcircle[red, fill=red]{2pt}
\tikzcircle[red, fill=red]{2pt}
& 
\tikzcircle[green, fill=green]{2pt}
\tikzcircle[green, fill=green]{2pt}
& & ++ & -\,- &  -\,-  \\

\cmidrule{2-2}
&\multicolumn{1}{l}{ $A_2$,$B_1$,$C_1$--\textit{Type7}} & 

\tikzcircle[green, fill=green]{2pt} 
&
\tikzcircle[green, fill=green]{2pt} 
& 
\tikzcircle[green, fill=green]{2pt} 
& 
 \tikzcircle[black, fill=white]{2pt} 
& 
\tikzcircle[red, fill=red]{2pt} 
\tikzcircle[red, fill=red]{2pt}
\tikzcircle[red, fill=red]{2pt} 
&
 \tikzcircle[black, fill=white]{2pt} 
&
 \tikzcircle[black, fill=white]{2pt} 
 & 
\tikzcircle[green, fill=green]{2pt}
 & 

& + & - & - \\

& \multicolumn{1}{l}{$A_2$,$B_1$,$C_{2/3}$--\textit{Type8/9}} &

\tikzcircle[green, fill=green]{2pt} 
\tikzcircle[green, fill=green]{2pt} 
&
\tikzcircle[green, fill=green]{2pt} 
\tikzcircle[green, fill=green]{2pt} 
& 
\tikzcircle[green, fill=green]{2pt} 
\tikzcircle[green, fill=green]{2pt} 
& 
\tikzcircle[red, fill=red]{2pt} 

& 
\tikzcircle[red, fill=red]{2pt}
\tikzcircle[red, fill=red]{2pt} 
\tikzcircle[red, fill=red]{2pt}
& 
\tikzcircle[green, fill=green]{2pt}
\tikzcircle[green, fill=green]{2pt}
&
\tikzcircle[red, fill=red]{2pt} 
\tikzcircle[red, fill=red]{2pt} 
&
\tikzcircle[green, fill=green]{2pt}
\tikzcircle[green, fill=green]{2pt}
&
& ++ & -\,- &  -\,-  \\

\cmidrule{2-2}
& \multicolumn{1}{l}{$A_2$,$B_2$,$C_1$--\textit{Type10}} & 

\tikzcircle[green, fill=green]{2pt} 
\tikzcircle[green, fill=green]{2pt} 
&
\tikzcircle[green, fill=green]{2pt} 
\tikzcircle[green, fill=green]{2pt} 
& 
\tikzcircle[green, fill=green]{2pt} 
\tikzcircle[green, fill=green]{2pt} 
& 
\tikzcircle[red, fill=red]{2pt} 
\tikzcircle[red, fill=red]{2pt} 

& 
\tikzcircle[red, fill=red]{2pt} 
\tikzcircle[red, fill=red]{2pt} 
\tikzcircle[red, fill=red]{2pt} 
&
\tikzcircle[green, fill=green]{2pt} 
\tikzcircle[green, fill=green]{2pt} 
&
\tikzcircle[red, fill=red]{2pt} 
\tikzcircle[red, fill=red]{2pt}
& 
\tikzcircle[green, fill=green]{2pt}
\tikzcircle[green, fill=green]{2pt}
& 
& ++ & -\,- &  -\,-  \\

& \multicolumn{1}{l}{$A_2$,$B_2$,$C_{2/3}$--\textit{Type11/12}} & 

\tikzcircle[green, fill=green]{2pt}
\tikzcircle[green, fill=green]{2pt}
\tikzcircle[green, fill=green]{2pt}
&
\tikzcircle[green, fill=green]{2pt}
\tikzcircle[green, fill=green]{2pt}
\tikzcircle[green, fill=green]{2pt}
& 
\tikzcircle[green, fill=green]{2pt}
\tikzcircle[green, fill=green]{2pt}
\tikzcircle[green, fill=green]{2pt} 
& 
\tikzcircle[red, fill=red]{2pt} 
\tikzcircle[red, fill=red]{2pt} 
\tikzcircle[red, fill=red]{2pt} 
& 
\tikzcircle[red, fill=red]{2pt} 
\tikzcircle[red, fill=red]{2pt} 
\tikzcircle[red, fill=red]{2pt} 
&
\tikzcircle[green, fill=green]{2pt}
\tikzcircle[green, fill=green]{2pt}
&
\tikzcircle[red, fill=red]{2pt} 
\tikzcircle[red, fill=red]{2pt}
& 
\tikzcircle[green, fill=green]{2pt}
\tikzcircle[green, fill=green]{2pt}
\tikzcircle[green, fill=green]{2pt}
& 

& +++ & -\,-\,- &  -\,-\,- \\ 
\bottomrule

\end{tabular} 
}
\begin{tablenotes}
 \item \quad \tikzcircle[black, fill=white]{2pt} Baseline, \tikzcircle[green, fill=green]{2pt} Property enhance, \tikzcircle[red, fill=red]{2pt} Property decrease; (Compared to \textit{Type1})
\end{tablenotes}
\end{table}

\smallskip
\textbf{\textit{A3: Hybrid Computation Architecture.}} This architecture shares the same workflow with $A1$ and $A2$. As discussed, on-chain computations are both time-consuming and money-consuming. Hybrid computation architecture solves this issue by bringing TTP with high computation capabilities. These TTPs process the data off-chain, and merely put a few computation tasks on the blockchain. In our example, the on-chain computation tasks may cover the token payment and NFT  the ownership transfer. This design improves user usability because off-chain computations have better performance and scalability.  The related architectural designs include \textcolor{brown}{{\textit{Type4-6}}}. However, it still assumes that the TTP who provides computation power is honest. A TTP in the real world may act maliciously or fail to provide computations due to hidden interests or being compromised. Again, such malicious computations are hard to be detected.

\smallskip
\textbf{\textit{A4: Hybrid Storage Architecture.}} This architecture stores the raw data off-chain while the data pointer is stored on the blockchain system. In our example, only NFT identifiers and payment information are stored on-chain, while the raw NFT data is stored off-chain. Correspondingly, in the real world, most NFT artists rely on centralized platforms like OpenSea \cite{opensea}, or Solanart \cite{solanaart} to store the raw NFT data. In this architecture, the choice of how to store off-chain data and how to reveal the contents of their off-chain data are left to users, which brings certain privacy and over-publicity issues. Meanwhile, the architecture improves the performance and saves the cost. Due to these benefits, the design has been widely adopted by some storage-intensive applications, such as blockchain-based streaming media, such as Theta \cite{theta}, Audius \cite{audius} and Livepeer \cite{livepeer}, which depends on the combination of on-chain storage and off-chain storage. The related architectural designs include \textcolor{teal}{\textit{Type10-12}}.

However, the solution is partially centralized with the risk of single point failures. In particular, the off-chain storage is controlled by a TTP. Any services built on top of this type can only be processed when a TTP is available, making the data access to relevant information becomes a privilege. Fortunately, some storage systems rely on replication to back up files for data integrity. For example, Storj \cite{wilkinson2014storj} utilizes the proof-of-redundancy mechanism, where every file is stored in at least three locations to avoid files being destroyed. The system operates on the Ethereum platform and stores the metadata in Satoshi format. Similarly, Sia~\cite{vorick2014sia} is a distributed storage system that relies on storage proofs. These proofs consist of a list of publicly verifiable root hashes from the submitted file and a fraction of the original file, and users can verify them easily from on-chain data. InterPlanetary File System (IPFS) \cite{filecoin}, with more complicated proof mechanisms, establishes a fully distributed peer-to-peer file system. Leveraging proof of time/space, IPFS ensures data integrity from both time and space.

\smallskip
{\textbf{Statistical Results.}} The investigation tries to answer the question \textit{which architectural design is the most prevailing?} Existing Web3 projects are designed based on different scenarios. Accordingly, their solutions target different components. We expand our research from mature systems (ranked in the market) as well as newly released whitepapers that claim to launch the projects. Based on our investigation (inevitably select examples in a small corpus from the entire project pool),  we find that most of the teams adopt the \textit{Type1} ($A_1,B_1,C_1$) design with a straightforward blockchain-based architecture. These projects either put focuses on merely one functionality, such as setting connections from Web2 to Web3 \cite{web3auth}, building decentralized identities \cite{idx,ceramic}, or implement the very basic infrastructure \cite{truffle,alchemy,ankr} that can better serve existing blockchain ecosystems. Above reasons made us to select \textit{Type1} as the baseline. For other types, we can observe that the options of external techniques are insufficient. For instance, if a project aims to move the storage off-chain, the options are limited to IPFS \cite{ipfs}, which has been implemented for years with comprehensive instructions and user guides. Affected by little attention to external technologies, other design types (even the sum of the rest ones) only occupy a small share of the entire picture.

\smallskip
\textbf{A Concise Summary.} As shown in Tab.\ref{tab-archievalua}, we provide detailed evaluations of each architectural design. Adding auxiliary techniques (agent, off-chain computation, off-chain storage) will impact the properties from different sides. Positively, performance, scalability, gas cost, and usability can get improved at different levels depending on how many layers have been modified. This is because external techniques can support much more volume of data and participating parties. From the view of single users, they can obtain better services due to faster transaction processing time and cheaper gas costs. In contrast, negatively, security, anonymity, and availability have decreased, compared to the baseline (\textit{Type1}), due to the in-transparency of off-chain processing procedures. Here, the singly applied agent at the client component will not affect the anonymity because either personal address or custodian addresses are equivalent towards adversaries. Similarly, confidentiality, security, and availability are majorly vulnerable at the computation and storage components, with little relation to the access component. Finally, for stakeholders, we can observe that Web3 users can obtain benefits that are consistent with previous reasons (better service). In opposite, both service providers and blockchain maintainers are disadvantageous since the managing costs accordingly increase.

\section{Extending to the Web3 Space}
\label{sec-discussion}
In this section, we extend our scope from architectures to the entire Web3 space. We present the Web3 impacts to current markets. Then, we point out the promising research directions as well as potential barriers on the road.

\subsection{Web3 Impacts}
\label{subsec-impact}

We first discuss the impacts of current Web3 projects. We approach it from two sides. The one is to estimate their influenced market value, while the other one is to see the scale of Web3 applications or services. We give details as follows.

\smallskip
\textbf{Value Estimation.} Supported by mainstream blockchain platforms, Web3 has gained an unexpected breakout. Measuring the quantitative impact is impractical due to its generality and vague bounder. No statistical data can be directly used by different institutions. But we still find indirect ways to reflect its influence. The first way is to investigate the usage of smart contract languages. This is because building a Web3 DApp requires defining logic and functions (e.g., ownership/transfer/connect) through these languages due to its close relation with user interfaces. The total value locked (TLV) by smart contract languages reaches up to maximally $193.14$ billion USD (as of Dec, 2021)\footnote{The TLV of $\$251.877$b is made up by $\$193.14$b from Solidy ($76.68\%$), $\$28.19$b from Rust ($11.19\%$), $\$27.56$b from Vyper $10.94\%$, $\$1.57$b from Ride ($0.62\%$), $\$1.07$b from Cairo ($0.42\%$), $\$193.47$m Bitcoin Script, $\$152.45$m from C\#, $\$3.69$m from Python. Among them, Solidity still occupies the largest proportion, indicating that Ethereum and EVM-compatible platforms play the most important role in the Web3 area. [Data source: \url{https://defillama.com/languages}].}. However, these languages only show a partial picture of the entire Web3 ecosystem, while many competitive programming languages are proposed following the same targets (cf. Appendix C). The real impact of Web3 will greatly outpace the stated data singly observed from languages. Another way to reflect the popularity of Web3 applications is based on fundamental index methodology \cite{web3index} that covers different types of underlying valuation figures such as users' paid fees on applications and individual votes for proposals. Fees, in this track, are different from transaction fees paid to miners in Ethereum. Instead, they are used to represent the cost that people are willing to spend on related decentralized services. For instance, users have spent $253,015$ USD (per $30$ days) on Arweave \cite{arweave} which is used to query or store data on the network. It provides an estimator of a network's value and activities by an in-time track from the view of the demand-side, which provides a distinguished way of investigating the status of the Web3 market. 

\smallskip
\textbf{Practical Development.} Besides its high values, Web3 DApps have promoted a rapid evolution towards a wide variety of scenarios. A promising paradigm shift is to move the centralized authority from trusted third parties (TTP) to distributed participants, which is also known as the Decentralized Autonomous Organization (DAO). DAO removes the formal management roles (e.g., delegated authorities) and physical entities (company/office)  \cite{fritsch2022analyzing}, by instead a suite of contracts residing on the Ethereum blockchain. DAOs have been adopted with many instances like Aragon \cite{aragon} and MetaCartel \cite{metacartel}. Another significant shift lies in their way of authenticating identities. Unlike asking for sensitive information through methods of \textit{email$+$password} or \textit{OAuth} in traditional networks, identities in Web3 applications are replaced by anonymous addresses, where the web establisher cannot obtain any useful knowledge of their users. The design further stimulates the widespread of decentralized identifiers (DIDs) \cite{w3cdid}. Users can fully control assets and metadata under their DIDs, and optionally release them to service providers by personal preferences. Moreover, Web3 also instimulates the prosperity of development tools and supplementary suites surrounding blockchain technologies. These tools cover APIs, statistic indicators (The Graph \cite{thegraph}), distributed storage (Sia \cite{sia}, Arweave \cite{arweave}), edge computing (Helium \cite{helium}), etc. They fulfill the blank left from previously isolated components, enabling them to seamlessly integrate together and well support each other. We further provide a look into Web3 applications that are stated in Tab.\ref{tab-projects} in Appendix D. Unfortunately, constrained by our bandwidth, we only select a small group of Web3 projects to demonstrate its wide adoption in this report.

\subsection{New Paradigm}
\label{subsec-oppo}
We then discuss the forthcoming wave of Web3 that goes beyond the initial use case like cryptocurrencies. Web3 can promote technical integration across different domains, ranging from establishing decentralized self-governance organizations to facilitating the progress of DeFi and Metaverse.

\textbf{Technical Integration.} Web3 is a novel approach to delivering internet architecture in a decentralized way. As discussed, Web3 covers every layer of the web architecture, from the front-end to the back-end. It indicates that technical integration may occur in multiple areas. For instance, Theta Labs' \cite{theta} decentralized video solution aims to stream video on Web3 via customized APIs. Audius \cite{audius} is a music sharing platform with the target to decrease the dependency on a record label. Radicle \cite{radicle} is an open-source and distributed platform for code collaboration. Arweave \cite{arweave} and Sia \cite{sia}, similar to IPFS \cite{ipfs}, are to establish a decentralized storage network that allows users to store data. Deeper Network \cite{deeper} intends to build a hardware-powered VPN ecosystem. All these attempts to modify, integrate and improve the existing infrastructure from centralized networks are promising, providing educational experiences on both success path and failure cases for the following developers. 

\textbf{Distributed Autonomous Organization (DAO).} Web3 will significantly promote the development of current distributed autonomous organizations. Utilizing a suite of smart contracts improves reliability, as no powerful authorities can break the rules. DAOs run on a flattened hierarchy where each participant has the right to vote on specific issues, similar to the way in a conventional executive board. All the processes, including decision-making, token issuing, option selecting and voting, are transparent due to the on-chain settings. This means anyone, either internally or externally, could audit the code, which greatly improves accountability and reliability and avoids misdirecting usage of funds collected from investors and users. To achieve decentralized governance, every project can issue its specified votes (or tokens) that stakeholders can put their preferences on. For instance, Yearn \cite{yearn} allows users to participate in decision-making and voting on proposals. Radicle \cite{radicle}, as a decentralized GitHub alternative, grants stakeholders the right to manage the project. Similarly, many DeFi protocols, including Uniswap \cite{uniswap}, The Graph \cite{thegraph}, SuperRare \cite{superrare}, and Audius \cite{audius}, enable ownership, participation, and governance via their issued tokens. All these votes and tokens require a Web3 DApp to get interacted with different DAO participants. Also, long-existing DAOs, such as MakerDAO \cite{makerdao}, have attracted many developers contributing to crypto-space and Web3 ecosystems. In this sense, DAO has extended the scope of decentralization, which was previously bound to machines, to a broader area that involves human beings. The shift of \textit{how to operate a digital organization} would pose a great impact on the future world, more than they are presented today.

\textbf{Self Governance.} Initially, software companies in Web1/Web2 obey the rule of protecting data, with a simple aim to involve more users for growth. But eventually, they have to start turning profits by selling or manipulating users' data, such as training AI models to make better recommendations. Users have no choice, facing the dilemma between privacy and convenience. In contrast, individuals in the Web3 space can hold as much personal data, which is more than ever before. Together with DiDs, users can freely browse the internet as well as perceive their data without compromising its privacy. Many solutions, like Ceramic \cite{ceramic} and IDX \cite{idx}, replace traditional authentication by allowing users to build self-sovereign identities. The Ethereum foundation makes much more progress by drafting an RFP \cite{rfpeth} for defining a formal specification. By controlling data and assets, an individual can even earn profits through incentive mechanisms. This is practically important for building a sustainable ecosystem that encourages users to continuously contribute to the required infrastructure.

\textbf{DeFi.} This is an intuitive application of Web3 as all the assets held by users are stored in their wallets. Performing financial-related activities, such as swapping different tokens at DEXes, loaning coins from exchanges, buying/selling crypto-assets from fiat to stablecoins, and designing derivatives (e.g., NFT \cite{wang2021non}, contracts, securities \cite{chen2022absnft}, share, etc.), becomes common and easier for single users since they do not need to register to any financial intermediaries. These activities make up the core of today's DeFi market \cite{werner2021sok}, which also attracts tremendous monetary investments involved in this field. Blockchain systems lay the foundation of DeFi protocols, guaranteeing the normal operations of every piece of logic, while Web3 paves the path to practical usage of these protocols, guiding users to participate in the games. Users only need to act as they behave in centralized markets, rather than having hard research to understand the differences between token standards or blockchain platforms.

\textbf{Metaverse.} Unlike DeFi or NFT specifying one specific direction, Metaverse is a general term that involves numerous technologies, with ambiguous targets to describe a virtual digital world for the incoming future. Intuitively, users in Metaverse will first interact with its front-end and then connect with the decentralized network supported by interoperable blockchain platforms. Web3 can cover all touchable applications that a user can reach, such as social networks, search engines (Brave \cite{brave}), galleries (Opensea \cite{opensea}), and marketplaces. A user can fully control their digital assets, identities, and data, browsing any sector of the Metaverse just like shopping at the store. Web3 can help to establish such a front-end environment with a pretty easy one-step connection for the participating users. In practice, many projects start to establish Metaverse from different aspects like IoT \cite{helium}, Games \cite{axieinfinity}, Markets \cite{superrare}\cite{opensea}, etc.

\subsection{Open Challenges}
\label{subsec-chall}

In this subsection, we point out the potential challenges from four folds, separately from the views of user-level (\textit{application}), system-level (\textit{blockchain}), markets (\textit{economy}), and social organizations (\textit{legality}).

\noindent\textbf{Application.} Decentralized applications built on blockchain-empowered systems are the first gate to individuals. Users have intuitive feelings towards these applications. We abstract three aspects that may affect the user experience.

\begin{itemize}
    \item[-] \textbf{Composability.} Web3-based solutions cover a wide range, either from the type of components, or the products for each component. Data transmitting in different products are inevitably isolated. The way to make data reused by different products at the same layer is urgently important. Standardized APIs may address the problem to some extent, where at least many DApps with similar designs can invoke the same APIs. Ceramic \cite{ceramic} tries to build applications with composable Web3 data and enable reusable data for multiple scenarios. However, making most back-end components composable is still a challenge. The barriers are whether composable components are compatible with each other and whether data can flow seamlessly across different components. For such reasons, the data need to be designed in the same structure for the re-usage by these components.
    

    \item[-] \textbf{Accessibility.} Web2 networks still cover most Internet users' activities, including social media, shopping, meeting, education, and payments. Users are accustomed to and enjoy their services due to their super convenience and easy accessibility. The lack of integration with modern web browsers and mobile applications limits the wide adoption of Web3 to end-users. An individual user will not change from the product that he used for a long to a new one. How to decrease the migration cost is a challenge. Moreover, interacting with Web3 applications also requires additional development, education, and software/hardware. This becomes a huge recognition difficulty for users and thus impedes its wide adoption.
    
    \item[-] \textbf{Data Recovery.} The private key is the most important secret of users when using blockchain platforms. The entire account of users relies on a credential that is represented in the form of a complex string or a series of secret recovery phrases. If a user loses his private key, he will never enter his account. The account becomes a dead account with all data, including assets, being locked. Methods to recover accounts, or at least the internal data, are an urgent requirement for applications in the Web3 space. Moreover, applications that adopt off-chain storage require more strict verification for data integrity because additional checks on whether the data in external storage matches on-chain hash values are needed. The data recovery in external pages is difficult due to the absence of traceability and accountability.
\end{itemize}

\smallskip
\noindent\textbf{Blockchain.} Blockchain systems are the most fundamental layer in each architectural design type, supporting upper-layer applications as well as connecting underlying storage. Limitations in blockchain systems will significantly constrain the development of the entire Web3 ecosystem.

\begin{itemize}
    \item[-] \textbf{Scalability.} Scalability represents the ability to process transactions along with the increased scale of networks, which is reflected by the increase or decrease in its performance. This is a long-term issue existing in blockchain systems due to intrinsic decentralization. Every major step needs time, such as data update, signature verification, etc. Among them, the consensus process has the most impacts since the more mining nodes joined the network, the more computational tasks were added. Web3 is closely integrated with underpinned blockchain systems, which, as a result, will face the similar issue as well. In most cases, transactions are slow on website interactions as state transitions need confirmation and propagation throughout the network. Underlying blockchain platforms, together with atop Web3 applications, require conquering the challenges of accommodating rapid growth and the demands of not compromising performance. Otherwise, users, with a high probability, may have a poor experience such as extremely slow loading speed of websites. 

    \item[-] \textbf{Interoperability.} Web3 applications, in the foreseen future, will be deployed on many blockchain platforms. This requires interoperable blockchain technologies to facilitate smooth state transitions, either homogeneously or heterogeneously. Creating interchangeable communication channels to connect isolated decentralized ledgers is still a challenge for the development. Current interoperability solutions, such as pegged sidechains \cite{deng2018research}, hash-locks \cite{herlihy2018atomic}, and trusted relays \cite{frauenthaler2020testimonium}, partially mitigate the problem: they enable transactions to cross over chains within their specified ecosystems (e.g., Polkadot and their para-chain slots \cite{wood2016polkadot}), but cannot make nature transactions operated across different ecosystems (Polkadot token on Ethereum). A prevailing method is to create the wrapped token that anchors the origin token with equal supplies as an alternative representation, such as the BTC coin existing in Ethereum with the representation of WBTC (an ERC20 format \cite{wbtc}). However, this makes more and more \textit{representative} tokens produced with no actual usage, increasing the waste and complexity.
    
    \item[-] \textbf{Contract Vulnerability.} The security of smart contracts \cite{atzei2017survey} will directly affect its connected Web3 applications. Smart contracts contain the business logic and operation specifications, which are key to the upper layer applications. Meanwhile, smart contracts can act as autonomous agents \cite{li2022smart}, decreasing the power of centralized service providers in combined protocols and securing on-chain data from users. Vulnerabilities existing in smart contracts, caused by design flaws or implementation errors, may result in thousands of monetary loss. Examples include the integer underflow/overflow attacks \cite{atzei2017survey}, DAO attacks \cite{daoattack}\cite{thedao} and Parity Multi-Sig Wallet attack \cite{parityattack}. Even worse, the scripting nature of contract programming languages (cf. Appendix C) and the non-updateable feature of smart contracts will significantly limit the growth of Web3 applications.  
\end{itemize}

\smallskip
\noindent\textbf{Economy.} As one of the major differences compared to Web1/Web2, users in the Web3 space can automatically obtain rewards according to their contributions. Users holding both digital assets and metadata in their wallets can freely trade them to earn profits. But there are still many concerns about disparate incentive mechanisms, operating costs, and technical debts.

\begin{itemize}
    \item[-] \textbf{Incentive.} Users adopt Web3 applications largely due to their considerable potential revenues. They can earn extra profits by conducting activities in Web3 networks: browsing websites (Basic Attention Token \cite{brave}), providing online storage (Arweave \cite{arweave}), playing games (Axie Infinity \cite{axieinfinity}), or selling self-created products (Opensea~\cite{opensea}). Even for the current stage, developers can obtain airdropped tokens from the project teams by testing their demos on the testnet. However, as more and more users participate in the game, the threshold of obtaining rewards becomes extremely high. Thus, designing a positive incentive mechanism that can cover as many users is crucial to attract new players joining the network.
    
    
    \item[-] \textbf{Cost.} The high gas fee has already become a major hurdle in using Ethereum. Sending an transaction will cost more than hundreds of dollars (executions on smart contracts cost more). As a result, applications with complicated computations cannot be deployed on-chain. This is the reason why many competitive public blockchains can co-exist in the market. Moreover, blockchain provides very constrained capabilities of storage, where many DApps can only put a small portion of code on-chain. A potential solution is to adopt more off-chain solutions \cite{gudgeon2020sok} that can take over the workload from on-chain to local servers without significantly breaking decentralization.
    
     \item[-] \textbf{Technical Debt.} Many Web3 applications are designed in a limited way to facilitate the entire software development cycle. This would cause many costs of additional rework in the latter processes, which is denoted as the \textit{technical debt} \cite{techdebt}. A suitable approach that aims for a long-term proposal can save many more costs because a bad design will accumulate \textit{interest} just as it is in monetary debt. Developing more in an inappropriate route can improve the difficulty of making updates. The development of a project falling into technical debt will reach a point where it is no longer possible to implement the protocol improvements that align with its initial vision. Developers should be sufficiently prudent when designing their products, even for the initial proof-of-concept implementations.
\end{itemize}

\smallskip
\noindent\textbf{Legality.} Governments and official organizations have a lot of concerns due to the rapid shift and huge change in the cryptocurrency world. Individual users are eager to have their non-infringeable rights and are afraid of an anarchy state. It is pretty difficult to reach the balance. Here, we stress the two discussed challenges in the view of the social side.

\begin{itemize}
    \item[-] \textbf{Governance.} Since anyone can launch Web3 projects, an increased number of applications will inevitably make the market segmented and unsupervised. Traditional authorities such as official organizations and governments become less influential than it is today. A large proportion of power is distributed to DAOs, which are made up of individuals. However, anyone who uses Web3 can arbitrarily establish DAOs without strict authentication or KYC (know your customer) steps. This causes instability in society \cite{kiayias2022sok}. For instance, (web3) tweets on decentralized social media platforms would be uncensorable without facing the risk of being punished for spreading rumors. Moreover, some illegal trades, such as porn or drugs, might be abused in unsupervised networks as no explicitly compulsive laws can forbid them. 
    
    \item[-] \textbf{Taxable Difficulty.} The intrinsic property of Web3 is to give the incentive back to users. Individuals who contribute more will obtain higher rewards. Their contributions can be in any form, such as deposited stakes (PoS-based chains \cite{kiayias2017ouroboros}), activities (many projects airdrop tokens based on this), attentions (e.g., BAT \cite{brave}, Cirus \cite{cirus}), or followers in traditional social media (Twitter, Discord, Facebook). If most of the current network users move to the Web3 world, collecting taxes from, at least, technology companies and Internet practitioners becomes extremely difficult. Even worse, hiding assets in accounts can help users directly make investments in financial markets and earn profits. The governments can never know what has happened and when this has occurred, nor for associated evidence used for taxes.
    
\end{itemize}

\section{Closing Results}
\label{sec-closing}

This section collects several popular questions that are frequently mentioned in communities. We accordingly answer them with our investigated results.

\noindent\hangindent 1em \textit{\textbf{What is Web3?}} Web3 is an umbrella term used to describe the next generation of internet services. It incorporates a wide range of components in the computer infrastructure. We have, in this work, investigated \textit{in the wild} solutions and found several common design patterns. We decouple Web3 services into three major components as in Sec.\ref{sec-archi}. With this, an application/service, in the context of our methodology, can be categorized into a Web3 space if \textit{at least one component is decentralized}. Based on the current view, an ideal Web3 application, operates all the services on-chain. However, this design can only support services with lightweight computations and storage. Supplementary services require external techniques.


\noindent\hangindent 1em \textit{\textbf{What is the cornerstone of Web3?}} Following the previous discussion,
a natural question is that: how to implement the decentralization of each component for building Web3. We review the identified twelve architectural designs and find an interesting result: every component centres around continuously operating blockchain platforms. The front-end wallet needs to connect with the on-chain operation, while the external storage also requires on-chain information for item searching and valid proofs. Therefore, the characterization through \textit{access}, \textit{computation} and \textit{storage} highly rely on the service of blockchain: \textit{access} to the blockchain, \textit{computation} in the blockchain, and \textit{storage} surrounding blockchain. These integrated components can establish a Web3 service thanks to the support from the blockchain. In this way, Web3 holds the property of decentralization, obtaining benefits from the blockchain.


\noindent\hangindent 1em \textit{\textbf{Is Web3 sufficiently decentralized?}} Web3 is not as decentralized as it appears to be \cite{werbach2018blockchain}, in which building a Web3 application still highly relies on a small corpus of companies. For instance,  centralized exchanges that can trade cryptocurrencies are majorly controlled by Binance \cite{binance} and Coinbase \cite{coinbase}, while wallets are affected by MetaMask \cite{metamask}, NFT products  \cite{wang2021non} by OpenSea \cite{opensea}, and stablecoins by Tether \cite{tether}. Meanwhile, a similar phenomenon also occurs in DeFi markets: Uniswap \cite{uniswap} has the most Total Value Locked (TVL); Dai \cite{dai} has dominated the decentralized stablecoins; Chainlink \cite{chainlink} has greatly outcompeted others in the price oracle. Moreover, many infrastructure-related blockchain companies that provide programming interfaces and development tools are concentred on a few companies like Alchemy \cite{alchemy}, Infura \cite{infura}, and Ankr \cite{ankr}. The high density of consolidation in the cryptocurrency field will inevitably result in a partial monopoly, where only a small group of oligarchs takes most of the resources. 

\noindent\hangindent 1em \textit{\textbf{Is Web3 secure?}}
The answer would be ``No". Web3 services rely on a suite of composable components to seamlessly work together. Problems in any single component may lead to fail. Come back to our NFT example, the raw data of the NFT may be erased due to the offline insecure storage by adversarial attacks or compromised managers. This is merely an example from the storage layer. In the real world, issues happening in other blockchain components will as well result in severe problems, such as transaction congestion due to poor scalability, slow confirmation caused by probabilistic consensus, or logical loopholes in smart contracts. The security of Web3 is closely related to correct operations of the entire system, which should be an all-level stack evaluation.

\noindent\hangindent 1em \textit{\textbf{Has Web3 addressed privacy worries?}} Web3 services cannot protect user privacy as it claimed. We give discussions from the website, which is the entry of a user to Web3 applications and services. Web-side privacy issues cover a set of design pitfalls and malicious attacks launched from the front-end, such as browsers or mobile applications. In many cases, an adversary may act like a normal operator in the system. For instance, the centralized authority can play the role of a tracker with abilities to record Ethereum addresses over a wide range of users, resulting in privacy violations with the help of its script embedding techniques \cite{winter2021s}. Once a user leaves evidence of using an address at a DeFi site, the malicious browser, which also holds your identities, can map to your physical entity. Besides, other traditional web attacks may also threaten the Web3 sites due to the shared engine.

\begin{center}
\tcbset{
        enhanced,
        colback=red!5!white,
        boxrule=0.1pt,
        colframe=red!75!black,
        fonttitle=\bfseries
       }
       \begin{tcolorbox}[
       title=Fact and Fiction - Truth to be Told of Web3,
        arc=5pt,outer arc=5pt,drop large lifted shadow]
       Based on our investigations from both technical and economic perspectives, Web3 cannot fully replace current web services. Instead, Web3 will still highly rely on the existing Internet infrastructure, including programming languages, communication protocols, agents, and storage. Fortunately, Web3 has reshaped conventional finance markets and facilitated individual self-governance. Users at least begin to pay much more attention to their digital assets covering both virtual data and cryptocurrencies.
       \end{tcolorbox}
\end{center}

\section{Conclusion}
\label{sec-conclu}

Web3 is an emerging concept prevailing in the entire crypto-world. Applications and services in the Web3 space, with non-custodial nature, allow users to control their data and obtain rewards. However, no clear definitions of such a \textit{buzzword} have formed. In this tech report, we fill the gap by investigating a large corpus of \textit{in the wild} projects titled with Web3. We dig into this topic by decoupling existing systems supporting blockchain-based Web3 services into separate core components, and accordingly discussing related features and properties for each potential combination. In total, we have identified twelve architectural design types and evaluated them with profound discussions. Based on our study, we present the new paradigms gained by Web3 and point out the design pitfalls. We further provide our answers to several questions from communities. To the best of our knowledge, this is the first research on Web3 from the view of blockchain.
\bibliographystyle{unsrt}
\bibliography{bib}

\begin{thebibliography}{100}

\bibitem{web3soul}
Eric~Glen Weyl, Puja Ohlhaver, and Vitalik Buterin.
\newblock Decentralized society: Finding web3's soul.
\newblock {\em Available at SSRN: \url{https://ssrn.com/abstract=4105763}},
  2022.

\bibitem{csnreport}
The web3 report q3 2021 (consensys).
\newblock {\em \url{https://consensys.net/reports/web3-report-q3-2021/}}, 2021.

\bibitem{woodweb3}
Wood Gavin.
\newblock Why we need web 3.0.
\newblock {\em
  \url{https://gavofyork.medium.com/why-we-need-web-3-0-5da4f2bf95ab}}, 2022.

\bibitem{dappradar}
Dappradar.
\newblock {\em \url{https://dappradar.com/}}, 2022.

\bibitem{mathdapp}
Mathdapp.
\newblock {\em \url{https://mathdapp.store/}}, 2022.

\bibitem{avax}
Avalanche network.
\newblock {\em \url{https://www.avax.network/}}, 2022.

\bibitem{solana}
Solana network.
\newblock {\em \url{https://solana.com/}}, 2022.

\bibitem{bsc}
Binance smart chain.
\newblock {\em \url{https://bscscan.com/}}, 2022.

\bibitem{rose}
Oasis network.
\newblock {\em \url{https://oasisprotocol.org/}}, 2022.

\bibitem{xu2021sok}
Jiahua Xu, Krzysztof Paruch, Simon Cousaert, and Yebo Feng.
\newblock Sok: Decentralized exchanges (dex) with automated market maker (amm)
  protocols.
\newblock {\em arXiv preprint arXiv:2103.12732}, 2021.

\bibitem{axieinfinity}
Axie infinity.
\newblock {\em \url{https://axieinfinity.com/}}, 2022.

\bibitem{wang2021non}
Qin Wang, Rujia Li, Qi~Wang, and Shiping Chen.
\newblock Non-fungible token (\text{NFT}): Overview, evaluation, opportunities
  and challenges.
\newblock {\em arXiv preprint arXiv:2105.07447}, 2021.

\bibitem{liu2021make}
Zhuotao Liu, Yangxi Xiang, Jian Shi, Peng Gao, Haoyu Wang, Xusheng Xiao, Bihan
  Wen, Qi~Li, and Yih-Chun Hu.
\newblock Make web3. 0 connected.
\newblock {\em IEEE Transactions on Dependable and Secure Computing (TDSC)},
  2021.

\bibitem{nakamoto2008bitcoin}
Satoshi Nakamoto.
\newblock Bitcoin: A peer-to-peer electronic cash system.
\newblock {\em Decentralized Business Review}, page 21260, 2008.

\bibitem{castro1999practical}
Miguel Castro, Barbara Liskov, et~al.
\newblock Practical byzantine fault tolerance.
\newblock In {\em The USENIX Symposium on Operating Systems Design and
  Implementation (OSDI)}, volume~99, pages 173--186, 1999.

\bibitem{garay2015bitcoin}
Juan Garay, Aggelos Kiayias, and Nikos Leonardos.
\newblock The bitcoin backbone protocol: Analysis and applications.
\newblock In {\em Annual International Conference on the Theory and
  Applications of Cryptographic Techniques (EUROCRYPT)}, pages 281--310.
  Springer, 2015.

\bibitem{garay2017bitcoin}
Juan Garay, Aggelos Kiayias, and Nikos Leonardos.
\newblock The bitcoin backbone protocol with chains of variable difficulty.
\newblock In {\em Annual International Cryptology Conference (CRYPTO)}, pages
  291--323. Springer, 2017.

\bibitem{wang2020sok}
Qin Wang et~al.
\newblock Sok: Diving into \text{DAG}-based blockchain systems.
\newblock {\em arXiv preprint arXiv:2012.06128}, 2020.

\bibitem{ipfs}
Ipfs: Filecoin.
\newblock {\em \url{https://filecoin.io/}}, 2022.

\bibitem{opensea}
Opensea.
\newblock {\em \url{https://opensea.io/}}, 2022.

\bibitem{chatzigiannis2021sok}
Panagiotis Chatzigiannis, Foteini Baldimtsi, and Konstantinos Chalkias.
\newblock Sok: Blockchain light clients.
\newblock {\em Cryptology ePrint Archive}, 2021.

\bibitem{karantias2020sok}
Kostis Karantias.
\newblock Sok: A taxonomy of cryptocurrency wallets.
\newblock {\em Cryptology ePrint Archive}, 2020.

\bibitem{swarm}
Swarm.
\newblock {\em \url{https://www.ethswarm.org/}}, 2022.

\bibitem{kazman1998architecture}
Rick Kazman, Mark Klein, Mario Barbacci, Tom Longstaff, Howard Lipson, and
  Jeromy Carriere.
\newblock The architecture tradeoff analysis method.
\newblock In {\em Proceedings. Fourth IEEE International Conference on
  Engineering of Complex Computer Systems (cat. no. 98ex193)}, pages 68--78.
  IEEE, 1998.

\bibitem{kalra2018zeus}
Sukrit Kalra, Seep Goel, Mohan Dhawan, and Subodh Sharma.
\newblock Zeus: analyzing safety of smart contracts.
\newblock In {\em The Network and Distributed System Security Symposium
  (NDSS)}, pages 1--12, 2018.

\bibitem{Li2022SoKTC}
Rujia Li et~al.
\newblock Sok: Tee-assisted confidential smart contract.
\newblock {\em The 22nd Privacy Enhancing Technologies Symposium (PETS)}, 3,
  2022.

\bibitem{bass2003software}
Len Bass, Paul Clements, and Rick Kazman.
\newblock {\em Software architecture in practice}.
\newblock Addison-Wesley Professional, 2003.

\bibitem{kiayias2017ouroboros}
Aggelos Kiayias, Alexander Russell, Bernardo David, and Roman Oliynykov.
\newblock Ouroboros: A provably secure proof-of-stake blockchain protocol.
\newblock In {\em Annual International Cryptology Conference (CRYPTO)}, pages
  357--388. Springer, 2017.

\bibitem{coinbase}
Coinbase.
\newblock {\em \url{https://www.coinbase.com/}}, 2022.

\bibitem{binance}
Binance.
\newblock {\em \url{https://www.binance.com/}}, 2022.

\bibitem{solanaart}
Solana art.
\newblock {\em \url{https://solanart.io/}}, 2022.

\bibitem{theta}
Theta network.
\newblock {\em \url{https://www.thetatoken.org/}}, 2022.

\bibitem{audius}
Audius.
\newblock {\em \url{https://audius.co/}}, 2022.

\bibitem{livepeer}
Livepeer.
\newblock {\em \url{https://livepeer.org/}}, 2022.

\bibitem{wilkinson2014storj}
Shawn Wilkinson, Tome Boshevski, Josh Brandoff, and Vitalik Buterin.
\newblock Storj a peer-to-peer cloud storage network.
\newblock 2014.

\bibitem{vorick2014sia}
David Vorick and Luke Champine.
\newblock Sia: Simple decentralized storage.
\newblock {\em Retrieved May}, 8, 2014.

\bibitem{filecoin}
Filecoin: a decentralized storage network.
\newblock {\em \url{https://filecoin.io/}}, 2022.

\bibitem{web3auth}
Web3auth.
\newblock {\em \url{https://web3auth.io/index.html}}, 2021.

\bibitem{idx}
Idx: Identity protocol for open applications.
\newblock {\em \url{https://idx.xyz/}}, 2022.

\bibitem{ceramic}
Ceramic network.
\newblock {\em \url{https://ceramic.network/}}, 2022.

\bibitem{truffle}
Truffle suite.
\newblock {\em \url{https://trufflesuite.com/}}, 2022.

\bibitem{alchemy}
Alchemy.
\newblock {\em \url{https://www.alchemy.com/}}, 2022.

\bibitem{ankr}
Ankr.
\newblock {\em \url{https://www.ankr.com/}}, 2022.

\bibitem{web3index}
The web3 index.
\newblock {\em \url{https://web3index.org/}}, 2022.

\bibitem{arweave}
Arweave.
\newblock {\em \url{https://www.arweave.org/}}, 2022.

\bibitem{fritsch2022analyzing}
Robin Fritsch, Marino M{\"u}ller, and Roger Wattenhofer.
\newblock Analyzing voting power in decentralized governance: Who controls
  daos?
\newblock {\em arXiv preprint arXiv:2204.01176}, 2022.

\bibitem{aragon}
Aragon.
\newblock {\em \url{https://aragon.org/}}, 2022.

\bibitem{metacartel}
Metacartel.
\newblock {\em \url{https://www.metacartel.org/}}, 2022.

\bibitem{w3cdid}
Reed Drummond, Sporny Manu, Sabadello Markus, Longley Dave, and Allen
  Christopher.
\newblock Decentralized identifiers (\textrm{DID}s) v1.0: Core architecture,
  data model, and representations.
\newblock {\em \url{https://www.w3.org/TR/did-core/}}, 2021.

\bibitem{thegraph}
The graph network.
\newblock {\em \url{https://thegraph.com/en/}}, 2022.

\bibitem{sia}
Sia network.
\newblock {\em \url{https://sia.tech/}}, 2022.

\bibitem{helium}
Helium network.
\newblock {\em \url{https://www.helium.com/}}, 2022.

\bibitem{radicle}
Radicle.
\newblock {\em \url{https://radicle.xyz/}}, 2022.

\bibitem{deeper}
Deeper network: The decentralized gateway and infrastructure for web3.0.
\newblock {\em \url{https://www.deeper.network/}}, 2022.

\bibitem{yearn}
Yearn.
\newblock {\em \url{https://yearn.finance/}}, 2022.

\bibitem{uniswap}
Uniswap.
\newblock {\em \url{https://uniswap.org/}}, 2022.

\bibitem{superrare}
Superrare.
\newblock {\em \url{https://superrare.com/}}, 2022.

\bibitem{makerdao}
Makerdao.
\newblock {\em \url{https://makerdao.com/}}, 2022.

\bibitem{rfpeth}
Request for proposals (rfp): Sign-in-with-ethereum.
\newblock {\em
  \url{https://notes.ethereum.org/@djrtwo/sign-in-with-ethereum-RFP}}, 2022.

\bibitem{chen2022absnft}
Hongyin Chen, Yukun Cheng, Xiaotie Deng, Wenhan Huang, and Linxuan Rong.
\newblock Absnft: Securitization and repurchase scheme for non-fungible tokens
  based on game theoretical analysis.
\newblock {\em arXiv preprint arXiv:2202.02199}, 2022.

\bibitem{werner2021sok}
Sam~M Werner, Daniel Perez, Lewis Gudgeon, Ariah Klages-Mundt, Dominik Harz,
  and William~J Knottenbelt.
\newblock Sok: Decentralized finance (defi).
\newblock {\em arXiv preprint arXiv:2101.08778}, 2021.

\bibitem{brave}
Brendan Eich.
\newblock Brave browser.
\newblock {\em \url{https://brave.com/}}, 2022.

\bibitem{deng2018research}
Liping Deng, Huan Chen, Jing Zeng, and Liang-Jie Zhang.
\newblock Research on cross-chain technology based on sidechain and
  hash-locking.
\newblock In {\em International Conference on Edge Computing}, pages 144--151.
  Springer, 2018.

\bibitem{herlihy2018atomic}
Maurice Herlihy.
\newblock Atomic cross-chain swaps.
\newblock In {\em Proceedings of the 2018 ACM Symposium on Principles of
  Distributed Computing (PODC)}, pages 245--254, 2018.

\bibitem{frauenthaler2020testimonium}
Philipp Frauenthaler, Marten Sigwart, Christof Spanring, and Stefan Schulte.
\newblock Testimonium: A cost-efficient blockchain relay.
\newblock {\em arXiv preprint arXiv:2002.12837}, 2020.

\bibitem{wood2016polkadot}
Gavin Wood.
\newblock Polkadot: Vision for a heterogeneous multi-chain framework.
\newblock {\em White Paper}, 2016.

\bibitem{wbtc}
Wrapped bitcoin.
\newblock {\em \url{https://wbtc.network/}}, 2022.

\bibitem{atzei2017survey}
Nicola Atzei, Massimo Bartoletti, and Tiziana Cimoli.
\newblock A survey of attacks on ethereum smart contracts (sok).
\newblock In {\em International Conference on Principles of Security and
  Trust}, pages 164--186. Springer, 2017.

\bibitem{li2022smart}
Rujia Li et~al.
\newblock How do smart contracts benefit security protocols?
\newblock {\em arXiv preprint arXiv:2202.08699}, 2022.

\bibitem{daoattack}
Understanding the dao attack.
\newblock {\em
  \url{https://www.coindesk.com/understanding-dao-hack-journalists/}}, 2016.

\bibitem{thedao}
Understanding the dao attack.
\newblock {\em
  \url{https://www.coindesk.com/learn/2016/06/25/understanding-the-dao-attack/}},
  2016.

\bibitem{parityattack}
An in-depth look at the parity multisig bug.
\newblock {\em
  \url{https://hackingdistributed.com/2017/07/22/deep-dive-parity-bug/}}, 2016.

\bibitem{gudgeon2020sok}
Lewis Gudgeon, Pedro Moreno-Sanchez, Stefanie Roos, Patrick McCorry, and Arthur
  Gervais.
\newblock Sok: Layer-two blockchain protocols.
\newblock In {\em International Conference on Financial Cryptography and Data
  Security (FC)}, pages 201--226. Springer, 2020.

\bibitem{techdebt}
Wiki: Technical debt.
\newblock {\em \url{https://www.wikiwand.com/en/Technical_debt}}, 2022.

\bibitem{kiayias2022sok}
Aggelos Kiayias and Philip Lazos.
\newblock Sok: Blockchain governance.
\newblock {\em arXiv preprint arXiv:2201.07188}, 2022.

\bibitem{cirus}
Cirus.
\newblock {\em \url{https://cirusfoundation.com/}}, 2022.

\bibitem{werbach2018blockchain}
Kevin Werbach.
\newblock {\em The blockchain and the new architecture of trust}.
\newblock Mit Press, 2018.

\bibitem{metamask}
Metamask.
\newblock {\em \url{https://metamask.io/}}, 2022.

\bibitem{tether}
Tether.
\newblock {\em \url{https://tether.to/}}, 2022.

\bibitem{dai}
Dai.
\newblock {\em \url{https://makerdao.com/}}, 2022.

\bibitem{chainlink}
Chainlink.
\newblock {\em \url{https://chain.link/}}, 2022.

\bibitem{infura}
Infura.
\newblock {\em \url{https://infura.io/}}, 2022.

\bibitem{winter2021s}
Philipp Winter, Anna~Harbluk Lorimer, Peter Snyder, and Benjamin Livshits.
\newblock What's in your wallet? privacy and security issues in web 3.0.
\newblock {\em arXiv preprint arXiv:2109.06836}, 2021.

\bibitem{ferdous2019search}
Md~Sadek Ferdous, Farida Chowdhury, et~al.
\newblock In search of self-sovereign identity leveraging blockchain
  technology.
\newblock {\em IEEE Access}, 7:103059--103079, 2019.

\bibitem{consensysdid}
Consensys: Blockchain use cases: Blockchain in digital identity.
\newblock {\em
  \url{https://consensys.net/blockchain-use-cases/digital-identity/#howddiworks
  }}, 2022.

\bibitem{aries}
Hyperledger aries.
\newblock {\em \url{https://www.hyperledger.org/use/aries}}, 2022.

\bibitem{ontid}
Ontology network, ont id.
\newblock {\em \url{https://ont.id/}}, 2022.

\bibitem{veramo}
uport - veramo.
\newblock {\em \url{https://veramo.io/}}, 2022.

\bibitem{wood2014ethereum}
Gavin Wood et~al.
\newblock Ethereum: A secure decentralised generalised transaction ledger.
\newblock {\em Ethereum project yellow paper}, 151(2014):1--32, 2014.

\bibitem{sompolinsky2015secure}
Yonatan Sompolinsky and Aviv Zohar.
\newblock Secure high-rate transaction processing in bitcoin.
\newblock In {\em International Conference on Financial Cryptography and Data
  Security (FC)}, pages 507--527. Springer, 2015.

\bibitem{hasan2005survey}
Ragib Hasan, Zahid Anwar, William Yurcik, Larry Brumbaugh, and Roy Campbell.
\newblock A survey of peer-to-peer storage techniques for distributed file
  systems.
\newblock In {\em International Conference on Information Technology: Coding
  and Computing (ITCC)}, volume~2, pages 205--213. IEEE, 2005.

\bibitem{benisi2020blockchain}
Nazanin~Zahed Benisi, Mehdi Aminian, and Bahman Javadi.
\newblock Blockchain-based decentralized storage networks: A survey.
\newblock {\em Journal of Network and Computer Applications}, 162:102656, 2020.

\bibitem{feng2019bug}
Xiaotao Feng et~al.
\newblock Bug searching in smart contract.
\newblock {\em arXiv preprint arXiv:1905.00799}, 2019.

\bibitem{benligiraydecentralized}
Burak Benligiray, Sa{\v{s}}a Milic, and Heikki V{\"a}nttinen.
\newblock Decentralized apis for web 3.0.
\newblock {\em \url{https://api3.org/}}.

\bibitem{lee2019using}
Wei-Meng Lee.
\newblock Using the web3. js apis.
\newblock In {\em Beginning ethereum smart contracts programming}, pages
  169--198. Springer, 2019.

\bibitem{androulaki2018hyperledger}
Elli Androulaki, Artem Barger, Vita Bortnikov, Christian Cachin, Konstantinos
  Christidis, Angelo De~Caro, David Enyeart, Christopher Ferris, Gennady
  Laventman, Yacov Manevich, et~al.
\newblock Hyperledger fabric: a distributed operating system for permissioned
  blockchains.
\newblock In {\em Proceedings of the thirteenth EuroSys Conference (EuroSys)},
  pages 1--15, 2018.

\bibitem{gorenflo2020fastfabric}
Christian Gorenflo, Stephen Lee, Lukasz Golab, and Srinivasan Keshav.
\newblock Fastfabric: Scaling hyperledger fabric to 20 000 transactions per
  second.
\newblock {\em International Journal of Network Management}, 30(5):e2099, 2020.

\bibitem{teal}
The algorand virtual machine (avm) and teal.
\newblock {\em
  \url{https://developer.algorand.org/docs/get-details/dapps/avm/teal/specification/}},
  2022.

\bibitem{gilad2017algorand}
Yossi Gilad, Rotem Hemo, Silvio Micali, Georgios Vlachos, et~al.
\newblock Algorand: Scaling byzantine agreements for cryptocurrencies.
\newblock In {\em Proceedings of the 26th Symposium on Operating Systems
  Principles (SOSP)}, pages 51--68, 2017.

\bibitem{pact}
The pact programming language.
\newblock {\em \url{https://github.com/kadena-io/pact}}, 2022.

\bibitem{kadena}
Kadena.
\newblock {\em \url{https://kadena.io/}}, 2022.

\bibitem{dune}
Dune network.
\newblock {\em \url{https://dune.network/}}, 2022.

\bibitem{sui}
Sui blockchain platform.
\newblock {\em \url{https://docs.sui.io/learn/about-sui}}, 2022.

\bibitem{ssc}
Ssc: Stellar smart contracts.
\newblock {\em
  \url{https://github.com/stellar-deprecated/docs/blob/master/guides/walkthroughs/stellar-smart-contracts.md}},
  2022.

\bibitem{solidity}
Solidity.
\newblock {\em \url{https://docs.soliditylang.org/en/v0.8.13/}}, 2022.

\bibitem{rust}
Ethereum for rust developers.
\newblock {\em
  \url{https://github.com/stellar-deprecated/docs/blob/master/guides/walkthroughs/stellar-smart-contracts.md}},
  2022.

\bibitem{javasp}
Ethereum for javascript developers.
\newblock {\em
  \url{https://ethereum.org/en/developers/docs/programming-languages/javascript/}},
  2022.

\bibitem{yul}
Yul docs.
\newblock {\em \url{https://docs.soliditylang.org/en/v0.5.3/yul.html}}, 2022.

\bibitem{vyper}
Vyper docs.
\newblock {\em
  \url{https://vyper.readthedocs.io/en/v0.1.0-beta.12/index.html}}, 2022.

\bibitem{lokhava2019fast}
Marta Lokhava, Giuliano Losa, David Mazi{\`e}res, Graydon Hoare, Nicolas Barry,
  Eli Gafni, Jonathan Jove, Rafa{\l} Malinowsky, and Jed McCaleb.
\newblock Fast and secure global payments with stellar.
\newblock In {\em Proceedings of the 27th ACM Symposium on Operating Systems
  Principles (SOSP)}, pages 80--96, 2019.

\bibitem{move}
Move docs.
\newblock {\em \url{https://docs.sui.io/build/move}}, 2022.

\bibitem{diem}
Diem.
\newblock {\em
  \url{https://developers.diem.com/docs/technical-papers/the-diem-blockchain-paper/}},
  2022.

\bibitem{oceanprotocol}
Ocean protocol.
\newblock {\em \url{https://oceanprotocol.com/}}, 2022.

\bibitem{syndicate}
Syndicate: Turn any wallet into a web3-native investing dao.
\newblock {\em \url{https://syndicate.io/}}, 2022.

\bibitem{utopialabs}
Utopia: Collaborative payroll and expense management for daos.
\newblock {\em \url{https://www.utopialabs.com/}}, 2022.

\bibitem{arbol}
Arbol.
\newblock {\em \url{https://www.arbolmarket.com/}}, 2022.

\bibitem{etherisc}
Etherisc.
\newblock {\em \url{https://etherisc.com/}}, 2022.

\bibitem{royal}
Royal.
\newblock {\em \url{https://royal.io/}}, 2022.

\bibitem{mirror}
Mirror: Create and connect your world on web3.
\newblock {\em \url{https://mirror.xyz/}}, 2022.

\bibitem{creaton}
Creaton.
\newblock {\em \url{https://app.creaton.io/#/}}, 2022.

\bibitem{gari}
Gari network.
\newblock {\em \url{https://gari.network/}}, 2022.

\bibitem{gitcoin}
Gitcoin: Build and fund the open web together.
\newblock {\em \url{https://gitcoin.co/}}, 2022.

\bibitem{linkdrop}
Linkdrop.
\newblock {\em \url{https://linkdrop.io/}}, 2022.

\bibitem{cointraffic}
Cointraffic.
\newblock {\em \url{https://cointraffic.io/}}, 2022.

\bibitem{manifold}
Manifold.
\newblock {\em \url{https://www.manifold.xyz/}}, 2022.

\bibitem{cloutart}
Clout.art.
\newblock {\em \url{https://clout.art/}}, 2022.

\bibitem{niftyink}
nifty.ink.
\newblock {\em \url{https://nifty.ink/}}, 2022.

\bibitem{steem}
steem.
\newblock {\em \url{https://steem.com/}}, 2022.

\bibitem{akasha}
Akasha.
\newblock {\em \url{https://akasha.org/}}, 2022.

\bibitem{cyberconnect}
Cyberconnect.
\newblock {\em \url{https://cyberconnect.me/}}, 2022.

\bibitem{coinvise}
Coinvise.
\newblock {\em \url{https://www.coinvise.co/}}, 2022.

\bibitem{rally}
Rally.
\newblock {\em \url{https://rally.io/ }}, 2022.

\bibitem{mynf}
Mynfteam.
\newblock {\em \url{https://www.mynf.team/}}, 2022.

\bibitem{status}
Status.
\newblock {\em \url{https://status.im/}}, 2022.

\bibitem{violet}
Violet.
\newblock {\em \url{https://violet.co/}}, 2022.

\bibitem{litentry}
Litentry.
\newblock {\em \url{https://www.litentry.com/}}, 2022.

\bibitem{spruceid}
Spruceid.
\newblock {\em \url{https://www.spruceid.com/spruceid}}, 2022.

\bibitem{crpcom}
Crypto.com.
\newblock {\em \url{https://crypto.com/}}, 2022.

\bibitem{zapper}
Zapper.
\newblock {\em \url{https://zapper.fi/}}, 2022.

\bibitem{rainbow}
Rainbow.
\newblock {\em \url{https://rainbow.me/}}, 2022.

\bibitem{zerion}
Zerion.
\newblock {\em \url{https://zerion.io/}}, 2022.

\bibitem{haskweb3}
Haskell web3 documentation.
\newblock {\em \url{https://hs-web3.readthedocs.io/en/latest/index.html}},
  2021.

\bibitem{anchorprotocol}
Anchor protocol.
\newblock {\em
  \url{https://docs.anchorprotocol.com/developers-earn/anchor-earn-sdk}}, 2022.

\bibitem{solana-web3.js}
Solana-web3.js.
\newblock {\em \url{https://solana-labs.github.io/solana-web3.js/}}, 2022.

\bibitem{civic}
Civic.
\newblock {\em \url{https://www.civic.com/}}, 2022.

\bibitem{fluxprotocol}
Flux protocol.
\newblock {\em \url{https://www.fluxprotocol.org/}}, 2022.

\bibitem{polygon}
Polygon.
\newblock {\em \url{https://polygon.technology/}}, 2022.

\bibitem{zksync}
zksync.
\newblock {\em \url{https://zksync.io/}}, 2022.

\bibitem{starkware}
Starkware.
\newblock {\em \url{https://starkware.co/starknet/}}, 2022.

\bibitem{zkrollups}
Zk-rollups.
\newblock {\em
  \url{https://docs.ethhub.io/ethereum-roadmap/layer-2-scaling/zk-rollups/}},
  2022.

\bibitem{optimism}
Optimism.
\newblock {\em \url{https://www.optimism.io/}}, 2022.

\bibitem{arbitrum}
Arbitrum.
\newblock {\em \url{https://arbitrum.io/}}, 2022.

\bibitem{optimisticrollups}
Optimistic rollups.
\newblock {\em
  \url{https://ethereum.org/en/developers/docs/scaling/optimistic-rollups/}},
  2022.

\bibitem{threaddb}
Threadb.
\newblock {\em \url{https://docs.textile.io/threads/}}, 2022.

\bibitem{gundb}
Gundb.
\newblock {\em \url{https://gun.eco/}}, 2022.

\bibitem{bittorrent}
Bittorrent.
\newblock {\em \url{https://www.bittorrent.com/}}, 2022.

\bibitem{hardhat}
Hardhat.
\newblock {\em \url{https://hardhat.org/}}, 2022.

\bibitem{blockchainfoundry}
Blockchain foundry.
\newblock {\em \url{https://blockchainfoundry.com/}}, 2022.

\bibitem{brownie}
Brownie.
\newblock {\em \url{https://eth-brownie.readthedocs.io/en/stable/}}, 2022.

\bibitem{settlemint}
Settlemint.
\newblock {\em \url{https://www.settlemint.com/}}, 2022.

\bibitem{deepdao}
Deepdao.
\newblock {\em \url{https://deepdao.io/organizations}}, 2022.

\bibitem{snapshot}
Snapshot.
\newblock {\em \url{https://snapshot.org/#/}}, 2022.

\bibitem{tally}
tally.
\newblock {\em \url{https://www.tally.xyz/}}, 2022.

\bibitem{nansen}
Nansen.
\newblock {\em \url{https://www.nansen.ai/}}, 2022.

\bibitem{tokenterminal}
Token terminal.
\newblock {\em \url{https://tokenterminal.com/}}, 2022.

\bibitem{messari}
Messari.
\newblock {\em \url{https://messari.io/}}, 2022.

\bibitem{theblock}
The block data dashboard.
\newblock {\em
  \url{https://www.theblockcrypto.com/data/nft-non-fungible-tokens/nft-overview}},
  2022.

\bibitem{neon}
Neon evm.
\newblock {\em \url{https://docs.neon-labs.org/docs/getting_started/}}, 2022.

\bibitem{ens}
Ens: Ethereum name service.
\newblock {\em \url{https://ens.domains/}}, 2022.

\bibitem{bonfida}
Bonfida: Solana name service.
\newblock {\em \url{https://naming.bonfida.org/#/}}, 2022.

\bibitem{api3}
Api3: The web api economy.
\newblock {\em \url{https://api3.org/}}, 2022.

\bibitem{pocket}
Pocket network.
\newblock {\em \url{https://www.pokt.network/}}, 2022.

\bibitem{datahub}
Datahub: The web 3 gateway.
\newblock {\em \url{https://datahub.figment.io/}}, 2022.

\bibitem{getblock}
Getblock: Superior node infrastructure for building dapps.
\newblock {\em \url{https://getblock.io/}}, 2022.

\bibitem{moralis}
Moralis.
\newblock {\em \url{https://moralis.io/}}, 2022.

\bibitem{quicknode}
Quicknode.
\newblock {\em \url{https://www.quicknode.com/}}, 2022.

\bibitem{figment}
Figment learn.
\newblock {\em \url{https://learn.figment.io/}}, 2022.

\end{thebibliography}

\section*{Appendix A. Web3 Primitives} 
\label{sec-bkg}
This section recalls several concepts used in Web3. We provide two main parts: one for the web-related knowledge, including its meaning, evolution, and architecture, while the others for basic primitives used to build a Web3 service. 

\subsection*{A.1 Web and Architecture}\label{subsec-web}
We first provide the common sense of so-called Web1/Web2, and abstract the backbone of their architectural designs. Based on that, we provide comparisons with Web3 from the perspective of its architecture.

\smallskip
\noindent\textbf{Web1/Web2.} The concepts of Web1/Web2 have become common knowledge for most internet users. We conservatively show their core properties from a high-level view during their evolution. The earliest version of the Web (Web1) is featured by its static sites, which consist of components such as images and text. Users access the targeted items by first finding a browser and then clicking what has been presented on the page. In this sense, Web1 is deemed as the \textit{read-only} web. Web2 extends Web1 by importing more complex designs (e.g. front/back-end architecture), enabling interactive actions from users by dynamic HTML. Users can both \textit{read and write} the content presented on sites, and can also upload or download files stored in databases. Users have their customized options of choosing which services are supported by providers (Facebook, Google, Amazon, etc.). This directly paves the way for various applications that require interactive web services, including marketplaces, user-generated content, and social media platforms. However, these centralized service providers gradually become the oligarch in their single fields because the big company has controlled huge amounts of data from users, some of which are even privacy-sensitive, like users' passwords or financial history. The advent of Web3 mitigates such issues by replacing the centralized back-end server with distributed ledgers. User can hold their personal accounts (containing digital assets) safely rather than relying on centralized banks. The services from each website, if interacting with users, have to be authenticated through the way of \textbf{\textit{connecting the wallet}} (cf. Sec.\ref{sec-protocol}). 

\smallskip
\noindent\textbf{Web Architecture.} We briefly summarize the potential architectural layouts and components of a web application. Building a typical web application relies on a client-server model. The client means the ends browsed by users through computers, or smartphones, while a web server serves the data and requests. Here, the webserver architecture is relatively complex that covers many fundamental aspects, including application tiers, operating systems (Windows, Linux, Solaris), platforms (.Net, LAMP), performance/quality of service (latency, throughput, low memory utilization), and physical capacity (computing power, storage, and memory). Receiving and responding to requests is the most basic action in successfully performing a web application. Firstly, a user visits a website by inputting a URL in the browser on the front-end. The browser parses the URL and sends the request to find the IP address via HTTPs. Then, the web server catches the request and processes it following the business logic (also called domain logic and application logic) in the back-end. The business logic manages the ways in which each piece of data is being accessed and determines the corresponding workflow, especially for each application. Last, the user receives the response on the web page sent from back-end servers.

\subsection*{A.2 Fundamental Components}\label{subsec-component}

Then, we summarise the basic components that are used to establish Web3 from the front user-end to the back server-end, covering \textit{light client} (wallet), \textit{VM-engined blockchains} (computation) and \textit{decentralised storage systems} (storage). Besides, we also introduce a close concept -- \textit{decentralized authentication}, which is important for individuals who physically connect themselves with online virtual identities (often in the form of anonymous addresses). 

\smallskip
\noindent\textbf{Decentralized Authentication.} Different from traditional ways of authentication, decentralized authentication removes, or at least weakens the dependency \cite{li2022smart} of trusted third parties (TTPs) during the procedures of verification and identifications. Each user accordingly has a unique identifier under the W3C commendations for decentralized identifiers (DIDs) and verifiable credentials (VC) \cite{w3cdid}. DID can be identified by the DID's controller, who might be a single person or an organization (also known as the self-sovereign identity \cite{ferdous2019search}). VCs are the proofs that follow an open standard for digital credentials, such as a passport or a license, or an ownership certificate of bank accounts. Each DID will be associated with a series of attestations generated by paired DIDs, usually in the form of VCs, to attest to its characteristics \cite{consensysdid}. Blockchain, in such cases, has two typical roles that either directly replace TTPs or assist existing ones by recording and managing the issuers' digital certificates \cite{li2022smart}. Several proof-of-concept projects have been proposed to highlight their targets to reshape DID files, including the open-source platforms developed by Hyperfabric Aries \cite{aries}, Ontology \cite{ontid}, uPort (now rebranded as Veramo \cite{veramo}), etc. Beyond that, if user privacy is an essential requirement, more complex security-related primitives (e.g., zero-knowledge proof, homomorphic encryption, or commitments) have to be introduced in the scheme.

\smallskip
\noindent\textbf{Light Client in Blockchain.} The term \textit{light} (equiv. \textit{lightweight}) \textit{client} shares a similar meaning of its usage for both Web2 and Web3: it merely receives the requests from users and forwards them to back-end servers (or blockchains) without participating in any logic processes. A slight difference is that a client in Web2 is often instantiated as a browser (covering both web browser, mobile browser, or an App), whereas in the context of Web3 or blockchain \cite{chatzigiannis2021sok}, it is typically represented as a \textit{wallet} \cite{karantias2020sok}, which is supported by locally running light nodes that connect to full nodes for information synchronization. A wallet interacts with the online blockchain system by sending a transaction to the \textit{Txpool} (transaction pool) and broadcasting them to peers via the gossip protocol \cite{nakamoto2008bitcoin}. Using such a client can pose more compatibility to resource-constrained environments such as different hardware devices, as well as reduce the costs of performing complex computations on-chain. The light client is an essential component in building Web3 applications, as it is the first entry to access the decentralized web environment. A shred of explicit evidence is that every Web3 page enforces users to  \textbf{\textit{connect the wallet}} (a button in the upper right position) when users want to conduct interactive actions on this website.

\smallskip
\noindent\textbf{VM-engined Blockchain.} The most significant difference between Web3 and previous web versions lies in the usage of blockchain. From the architecture perspective, blockchain replaces the traditional centralized back-end servers with distributed ledger systems like Ethereum \cite{wood2014ethereum}. A blockchain-based system takes over the tasks of processing business logic and responding to users (under Web2 semantics). This depends on well-functioned on-chain \textit{virtual machine} (VM) that enables state transitions. VM, in the context of blockchain, can be equivalently regarded as smart contracts that are automatically operated following the coded rules. The rules contain the logic requested from clients. Smart contracts, in this sense, play an essential role in enabling Web3 and DApps. Meanwhile, to ensure consistency across distributed nodes, a consensus protocol is required in each specific blockchain system. The consensus mechanisms solve the conflicts and maintain the chain stability by initiating a set of predefined principles (\textit{the longest-chain rule} in PoW \cite{nakamoto2008bitcoin}, \textit{the weightiest-chain} rule in GHOST \cite{sompolinsky2015secure}, and more \cite{wang2020sok}). 
Further, all these nodes mutually communicate in a P2P network. These components make up a typical blockchain architecture, which acts just like a fully functional back-end server from an external view. 

\smallskip
\noindent\textbf{Distributed Storage towards Blockchain.} File storage systems in distributed networks are fundamental for sharing and storing sensitive content across different nodes. Two major ways of distributed storage are either increasing data availability (replication) or reducing data loss (erasure code). Adopting a distributed file system can obtain benefits including fault tolerance, scalability, availability, and performance \cite{hasan2005survey}. In traditional ways, a lot of servers (on a scale of hundreds/thousands) have to cooperate to execute tasks requested from clients and applications, including service providers like Hadoop File System (HDFS), CernVM File System (CVFMS), and Andrew File System (AFS). Even though these providers deploy many machines in different areas, services are still controlled by providers, resulting in partial centralization. Differing from them, blockchain-based storage systems remove trusted central parties, which are theoretically more secure than centralized storage \cite{benisi2020blockchain}.

\begin{figure}[h!]
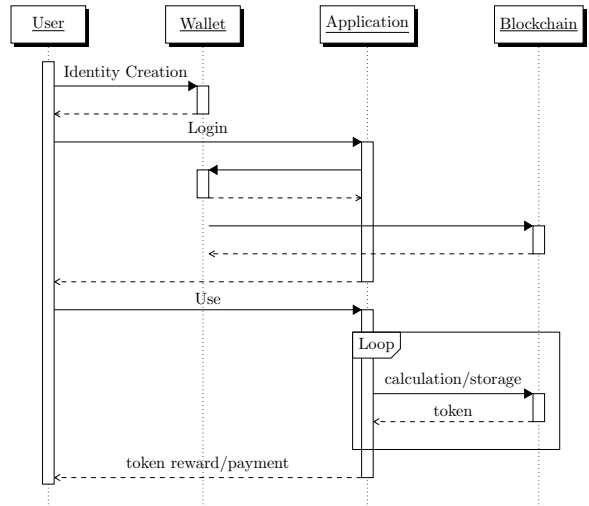

\centering
 \resizebox{0.67\linewidth}{!}{
\begin{sequencediagram}
\newthread[white]{u}{User}
\newinst[1.7]{w}{Wallet}
\newinst[1.7]{a}{Application}
\newinst[1.7]{b}{Blockchain}

\begin{call}{u}{Identity Creation}{w}{}
\end{call}
\begin{call}{u}{Login}{a}{}
\begin{call}{a}{}{w}{}
\end{call}
\begin{call}{w}{}{b}{}
\end{call}
\end{call}

\begin{call}{u}{Use}{a}{token reward/payment}

\begin{sdblock}{Loop}{}
\begin{call}{a}{calculation/storage}{b}{token}
\end{call}
\end{sdblock}

\end{call}

\end{sequencediagram}
}
\caption{Workflow of A Web3 System}
\label{fig-workflow}
\end{figure}

Blockchain storage mainly relies on redundant replications for data security in case of loss. Storj \cite{wilkinson2014storj} operates on Ethereum and replicates the metadata in multiple locations. Sia \cite{vorick2014sia}, as a distributed storage system, uploads a list of publicly verifiable root hashes of submitted files to get verification on-chain. InterPlanetary File System (IPFS) \cite{filecoin}, a peer-to-peer file system, leverage the proof-of-spacetime and proof-of-replication to guarantee that (i) the data is being stored during a specific duration of time and (ii) the data has been distributed in multiple hardware, rather than a single physical storage location. Based on such investigations, we abstract the distributed storage components, which can take most of the workloads when on-chain capacity is not enough.

\subsection*{A3.Web3 Workflow}\label{subsec-workflow}
In a Web3 system (cf. Fig.\ref{fig-workflow}), user's data is processed and stored in a decentralized and community-driven network, using open protocols instead of centralized TTP. An important feature of Web3 lies in its instant rewards, enabling users to obtain a fair share of revenues when they contribute to the network. 

\section*{Appendix B. Token Standards}
\begin{figure}[!] 
    \centering
    \includegraphics[width=0.8\linewidth]{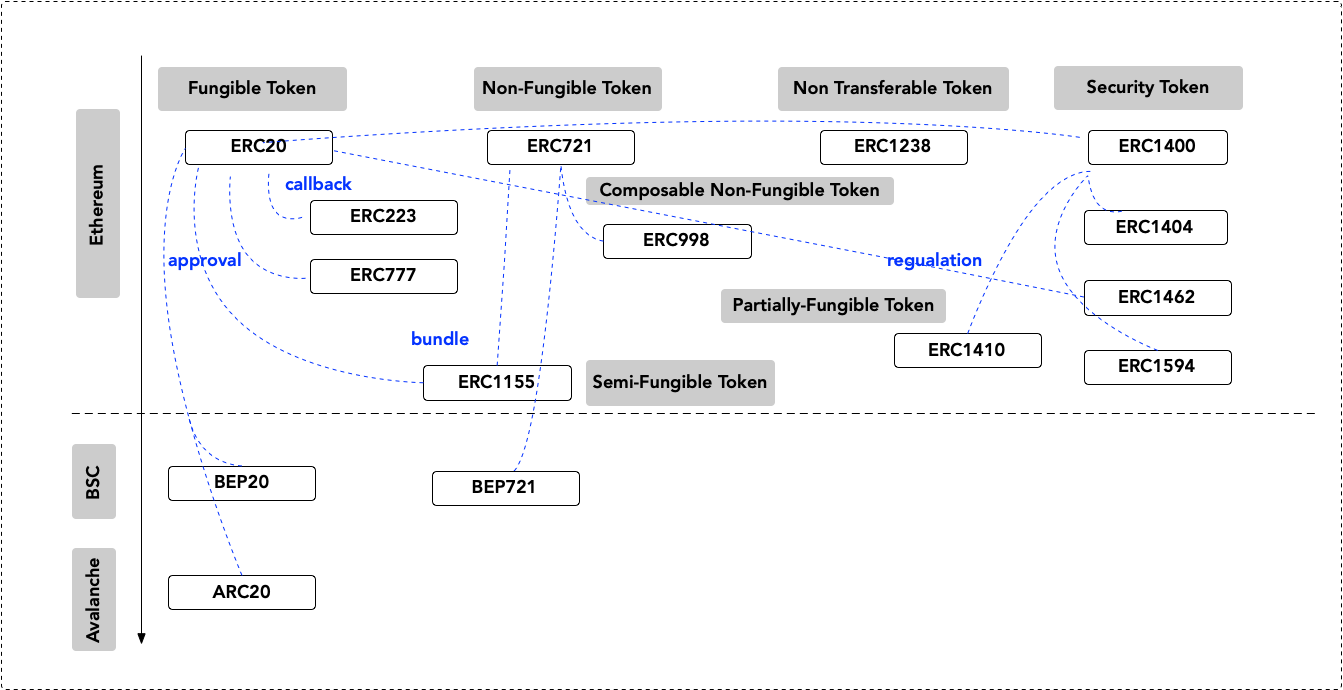}
   \caption{Relationship between Different Token Standards}
   \label{fig-tokenrelation}
\end{figure}

Tokens in Web3 play an essential role in incentivizing users and developers. The token standard, as the subsidiary of the smart contract standard, defines the methods to create, deploy and issue new tokens. We have investigated existing token standards from different blockchain platforms and summarised them in Tab.\ref{tab-tokenstandard}. Most token standards are issued through Ethereum, which has the biggest and most mature on-chain virtual machine and smart contract. The standards in the table are not separate, and many of them have close relations. As a result, these standards set the baseline of the entire ecosystem, even for the token standards in other competitive blockchain systems. For instance, Binance smart chain and Avalanche follow very similar principles. We give a brief guide in Fig.\ref{fig-tokenrelation}.

\begin{table}[!hbtp]
 \caption{Summary of Token Standards} 
 \label{tab-tokenstandard}
  \centering
 \resizebox{1\linewidth}{!}{
 \begin{tabular}{lcclrr}
    \toprule
    \multicolumn{1}{c}{\textbf{Standard}\quad\quad}   & \multicolumn{1}{c}{\textbf{Date}\quad\quad}  
     & \multicolumn{1}{c}{\textbf{Platform}\quad\quad}  & \multicolumn{1}{c}{\quad\textbf{Feature}\quad} &
    \multicolumn{1}{c}{\quad\quad\textbf{Application}\quad\quad} & \multicolumn{1}{c}{\quad\textbf{Notes}\quad} \\
    \midrule
   ERC20  & 2015  & Ethereum & Token API / Fungible Token & Vote/ICO & \\
   ERC721 & 2018 & Ethereum  & Non-Fungible Token & Artwork/IP &  \\
   ERC777 & 2018 & Ethereum  & Token Approval & & Improving ERC20 \\
   ERC1155 & 2018 & Ethereum  & Semi-Fungible Token & Game & Bundling of ERC20  \\
   ERC223 & 2017 & Ethereum  & Token Recovery  & &  \\
   ERC998 & 2018 & Ethereum  & Composable Non-Fungible Token &  Game/Ownership &  \\
   ERC1238 & 2018 & Ethereum  & Non-Transferrable Non-Fungible Token & Badge & \\
   ERC1594 & 2018 & Ethereum  & Core Security Token Standard & Securities (Financial) & \\
   ERC1400 & 2018 & Ethereum  & Security Token Standard &  Securities  &  \\
   ERC1404 & 2018 & Ethereum & Simple Restricted Token Standard & Securities & \\
   ERC1410 & 2018 & Ethereum & Partially Fungible Token Standard &  \\
   ERC1462 & 2018 & Ethereum  & Base Security Token &  Securities & \\
   \midrule
   BEP20 & 2020 & Binance  & Fungible Token  & Vote/Wrap Token &  \\
   BEP721 & 2020 & Binance  & Non-Fungible Token & IP Products & \\
   ARC721 & 2021 & Avalanche  & Fungible Token & Wrap Other Tokens  & \\
   
   \bottomrule
  \end{tabular}
 }
\end{table}

The most important token standard in the crypto-world is ERC20. It introduces the concept of fungible tokens and defines the software interface and token APIs. An ERC20 token is different from a chain-based ETH token because ERC20 tokens run based on smart contracts. By deploying a token-issue contract, everyone can create their tokens without initiating a separate blockchain. Such a design can be used as a variant type of \textit{Initial Public Offering} (IPO), where any team can launch a fundraising activity, denoted as the \textit{Initial Coin Offering} (ICO). This reduces the complexity of implementation and increases the liquidity across different tokens in the Ethereum ecosystem. ERC20 interfaces contain six major methods, namely, \texttt{totalSupply}, \texttt{balanceOf}, \texttt{transfer}, \texttt{transferFrom}, \texttt{approve} and \texttt{allowance} and two events: \textit{transfer} and \textit{approval}. These methods lay the foundation of all following standards. Another essential token standard is ERC721, a standard for the non-fungible token (NFT). Tokens in this type are distinguished where the pair of \textit{contract address} and \textit{tokenId} (in the form of a \texttt{uint256} variable) must be globally unique. The exact value of an NFT, reflected in the financial market, may vary in a huge range due to its rarity, age, or attention. Based on such attributes, ERC721 tokens are suitable to offer IP-related products \cite{wang2021non} that cover collectible items (images, songs, or books), tickets (for events, lotteries, or concerts), access keys, etc.

Beyond the wide adoption of ERC20 and ERC721 standards, other standards also contribute to communities by extending the functionalities and availability. ERC223 improves ERC20 (means remaining backward compatible with ERC20) by adding the \texttt{tokenFallBack}. In case of the monetary loss of sending tokens to a contract address, ERC223 can recover the missing tokens through the new function. ERC777 extends ERC20 by introducing a method to interact with the token contract. Users can build a mixer contract for additional functionalities like setting an emergency recovery function in case of the loss of private keys. ERC1155 is a semi-non fungible token that combines the features from both ERC20 and ERC721. The contact can manage multiple token types (e.g., NFT) where each type may contain a set of fungible tokens. In this way, it can process transaction bundles with high efficiency. ERC998, as an extension of ERC721, is a composable non-fungible token standard. It enables the integration of different ERC721 and ERC20 tokens, in which one can hold another non-fungible token at the same time. These combined NFTs are connected by ownership and organized like a tree. Also, there are several types of standards used for specific scenarios. ERC1238, non-transferrable non-fungible token (NTT), is designed with attributes of non-transferability. This can be used in recording users' reputation and experience (quantitative) or granting badges and titles (qualitative). ERC1410 adds an extra layer of granular transparency that can be used for the investigation of contract behaviors in different partitions. Another series of ERC20-extended standards, including ERC1400, ERC1404, ERC1461, and ERC1594, are designed for the security tokens in financial markets and official governments. Besides, we have noticed that BEP20 and BEP721 on Binance smart chain, and ARC20 on Avalanche, inherit attributes and functions in ERC standards, decreasing the cost of absorbing new bits of knowledge for newcomers.

\section*{Appendix C. Languages in VM-empowered Blockchains}

\begin{figure}
    \begin{minipage}[!h]{0.5\linewidth}
    \centering
    \includegraphics[height = 4.3cm]{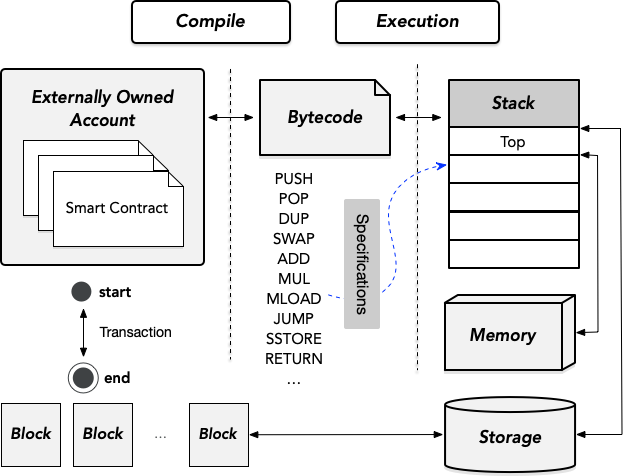}
    \end{minipage}
    \begin{minipage}[!h]{0.5\linewidth}
    \centering
    \resizebox{\linewidth}{!}{
    \begin{tabular}{l|r} 
    \toprule
    \multicolumn{1}{c}{\quad\textbf{Languages}\quad}   & \multicolumn{1}{c}{\quad\textbf{Blockchain}\quad}  \\
    \midrule
   Solidity \cite{solidity} & Ethereum, BSC, Avalanche    \\
   Rust \cite{rust} & Solana, Polkadot, Near  \\ 
   JavaScript \cite{javasp} & Hyperledger, FastFabric  \\ 
   Yul \cite{yul} & EVM-compatible  \\ 
   Vyper \cite{vyper} & EVM-compatible   \\ 
   TEAL \cite{teal} & Algorand  \\ 
   SSC \cite{ssc} & Stellar \cite{lokhava2019fast}   \\ 
   Pact \cite{pact}  & Kadena \cite{kadena}    \\ 
   Liquidity   & Dune Network \cite{dune}   \\  
   Move \cite{move} & Sui\cite{sui}, Diem\cite{diem}  \\
   \bottomrule
    \end{tabular}
    }
    \end{minipage}
    \caption{VM-empowered Blockchains Operations and Programming Languages}
    \label{fig-vmblockchian}
\end{figure}

Deploying the smart contract on-chain that contains the business logic is an essential procedure for Web3 services. As plenty of literature has introduced the operating mechanisms of blockchain virtual machines (see the skeleton \cite{feng2019bug} in the \textit{left figure} in Fig.\ref{fig-vmblockchian}) and security considerations of smart contracts, we skip these parts and put focus on filling the blank of programming languages in VM-supported blockchain platforms (the \textit{right table} in Fig.\ref{fig-vmblockchian}), which is rarely discussed but of great importance in the Web3 ecosystem. Smart contract programming languages enable writing programs and logic according to the requirements of users. These languages are typically targeted toward primary developers, requiring them to be sufficiently friendly. A contract written in such languages, then, is compiled to a bottom language such as binary codes to allow machines to execute. Corresponding actions are operated on-chain under the guide of specifications. We review existing smart contract programming languages to provide a guideline for developers with the aim to build Web3 applications and services.

\textit{Solidity} is an undoubtedly first-ranked language used in current blockchain systems. Benefits from the influence of Ethereum, Solidity has been widely adopted by most EVM-compatible blockchains such as Binance smart chain (BSC), Avalanche (c-chain), and Oasis Network (ParaTimes). The language is an object-oriented and statically-typed language that brings many similar designs from matured programming languages such as C++ and Python. For instance, Solidity supports inheritance, libraries, and complex user-defined types. Meanwhile, the language is Turing-complete which enables multi-functional developments. Users can customize their methods to realize different functionalities. \textit{Vyper} is a contract-oriented language that aims to improve the security of Solidity. It has many features that are designed for smart contracts, such as event notifiers for listeners, custom global variables, and global constants. The language cannot support complex features of inheritance, function overloading, infinite-length loops, and recursive calling to make it simple enough. It can be used in EVM-compatible systems without any barriers. Similarly, \textit{Yul} is designed to be an intermediate programming language that can be compiled to the format of bytecode used for the adjustment of different backends. The Solidity compiler has an experimental implementation that uses Yul as an intermediate language. Yul is used in stand-alone mode and for inline assembly inside Solidity. The language supports both EVM and ewasm (Ethereum flavored WebAssembly).

\textit{Rust} is a low-level statically-typed language, with features of being fast and memory-efficient. Also, no garbage collector exists in the language, meaning that the incidents caused by the language will happen with a negligible possibility. Due to its high efficiency, many blockchain systems have started to utilize Rust as their smart contract languages, including Solana, Polkadot, and Near Blockchain. \textit{JavaScript} is a general-purpose programming language, as well as an entry-level language that is adopted by most blockchains to create a JavaScript wrapper or library \cite{benligiraydecentralized,lee2019using}. Hyperledger Fabric \cite{androulaki2018hyperledger} and FastFabric \cite{gorenflo2020fastfabric} enable users to create a smart contract with several languages, including JavaScript (Node.js). Besides these mainstream languages, many platforms propose customized languages. TEAL \cite{teal} is an assembly language syntax used in Algorand \cite{gilad2017algorand} to specify programs. The language will be converted to the bytecode that can be recognized by its interpreter. Pact \cite{pact} is immutable, Turing-incomplete language used in Kadena \cite{kadena}. It uses a declarative approach over complex control flow, which makes bugs easier to be detected. With the same scope, Dune Network \cite{dune}, Sui \cite{sui} and Stellar \cite{ssc} propose their customized languages called \textit{Liquidity}, \textit{Move} and \textit{SSC}, respectively.

\section*{Appendix D. Web3 Projects}

\begin{table}[!hbtp]
 \caption{A Collection of Web3 Projects} 
 \label{tab-projects}
  \centering
 \resizebox{\linewidth}{!}{
 \begin{tabular}{p{1.5cm}p{3cm}lr}
 
    \toprule
    & \textbf{Functionality} &
    \multicolumn{1}{c}{\textbf{Project}} &
    \multicolumn{1}{c}{\textbf{Feature}}   
      \\
   
   \midrule
   \multirow{14}{*}{\rotatebox[origin=c]{70}{\textbf{Application}}} 
   & \multirow{2}{*}{\textbf{\makecell[l]{DAO}}} 
   &  MakerDAO \cite{makerdao}, OceanDAO \cite{oceanprotocol} &  \multirow{2}{*}{Voting-based Rights}  \\
   &  & Syndicate \cite{syndicate}, Utopia \cite{utopialabs}  \\
     
   \cmidrule{2-3} 
   & \multirow{12}{*}{\textbf{DApps}} 
   & Arbol \cite{arbol},  Etherisc \cite{etherisc} & Parametric Insurance    \\ 
   & & Theta \cite{theta},  Livepeer \cite{li2022smart} & Streaming Media (Video)   \\
   & & Royal \cite{royal}, Audius \cite{audius} & Streaming Media (Audio)\\ 
   & & Mirror \cite{mirror}, Creaton \cite{creaton}, Gari \cite{gari} & Content Management \\
   & & Radicle \cite{radicle}, Gitcoin \cite{gitcoin}, Yearn \cite{yearn}  & Code Repository  \\ 
   & & Linkdrop \cite{linkdrop}, Cointraffic \cite{cointraffic}  & User Acquisition \\
   & & Manifold \cite{manifold}, CloutArt \cite{cloutart}, NiftyInk \cite{niftyink} & NFT Platform \\
   & & Steem \cite{steem}, Akasha \cite{akasha} & Social Network \\
   & & Cyberconnect \cite{cyberconnect}, Coinvise \cite{coinvise}, Rally \cite{rally} & SocialFi \\
   & & Axie Infinity \cite{axieinfinity} & GameFi \\
   & & MyNFTeam \cite{mynf}  & Employment Platform \\
   & & Status \cite{status} & Messaging   \\
    
   \cmidrule{1-3} 
   \multirow{9}{*}{\rotatebox[origin=c]{70}{\textbf{Access}}}  
    & \multirow{2}{*}{\textbf{Identity}} 
    & IDX \cite{idx}, Violet \cite{violet}, Litentry \cite{litentry} &   \multirow{2}{*}{(W3C)DID-compatible}     \\
    & & Ceramic \cite{idx}, Spruce ID  \cite{spruceid} &   \\

   \cmidrule{2-3} 
   & \multirow{2}{*}{\textbf{Wallet}} 
   & MetaMask \cite{metamask}, Crypto.com  \cite{crpcom}  &  \\
   & & Zapper \cite{zapper}, Rainbow \cite{rainbow}, Zerion \cite{zerion} & \\
   
   \cmidrule{2-3} 
   & \multirow{2}{*}{\textbf{Client}} 
   & Web3.js, Ethers.js, Haskell \cite{haskweb3}  & Ethereum \\
   & & Anchor \cite{anchorprotocol}, @solana/web3.js \cite{solana-web3.js} & Solana \\
   
   \cmidrule{2-3} 
   & \multirow{1}{*}{\textbf{Authentication}} 
   & Web3auth \cite{web3auth}, Civic \cite{civic} & Link (Web2)Account with Address \\
  
   \cmidrule{2-3}  
   & \multirow{1}{*}{\textbf{Browser}}  
   &  Basic Attention Token \cite{brave} &     \\
   
   \cmidrule{1-3} 
   \multirow{11}{*}{\rotatebox[origin=c]{70}{\textbf{Computation}}}  
   & \multirow{2}{*}{\textbf{Oracle}} 
   & Chainlink \cite{chainlink}  &  \multirow{2}{*}{Capture External Data}    \\
   & & Flux \cite{fluxprotocol}  &      \\
   
   \cmidrule{2-3}      
   & \multirow{1}{*}{\textbf{Indexing}}  
   & The Graph \cite{thegraph} &    \\ 
   
   \cmidrule{2-3} 
   & \multirow{3}{*}{\textbf{\makecell[l]{Layer-1 \\ Blockchain}}} 
    &  Ethereum, BSC, Avalanche, Celo & EVM-compatible Chains \\
    &  & Cosmos, Polkadot, Solana &   \multirow{2}{*}{Competitive Chains} \\
    &  & Near, Celo, Aurora, Fantom, Tezos   \\
    
   \cmidrule{2-3} 
   & \multirow{3}{*}{\textbf{\makecell[l]{Layer-2 \\ Blockchain}}}
   & Ploygon \cite{polygon}  & Sidechain \\
   & & ZkSync \cite{zksync}, Starknet \cite{starkware}   & ZKrollups \cite{zkrollups}  \\
   & & Optimism \cite{optimism}, Arbitrum \cite{arbitrum}  & Optimistic Rollups \cite{optimisticrollups} \\

   \cmidrule{1-3} 
   \multirow{4}{*}{\rotatebox[origin=c]{70}{\textbf{Storage}}}  
   & \multirow{2}{*}{\textbf{Off-chain Data}} 
   & Ceramic Network \cite{ceramic}  &    \\   
   & & ThreadDB \cite{threaddb}, GunDB \cite{gundb}  &  \\

   \cmidrule{2-3} 
   & \multirow{2}{*}{\textbf{File Storage}}  
   & IPFS \cite{ipfs}, BitTorrent \cite{bittorrent}   &   \multirow{2}{*}{Distributed File Storage}  \\ 
   & & Arweave \cite{arweave}, Siacoin \cite{sia}   &  \\

  \cmidrule{1-3}   
  \multirow{17}{*}{\rotatebox[origin=c]{70}{\textbf{Supporting Tech}}}  
   & \multirow{5}{*}{\textbf{\makecell[l]{Developing \\ Tool Set}}}  
   & Truffle \cite{truffle}, Hardhat \cite{hardhat} &  JavaScript \\
   & & Foundry \cite{blockchainfoundry} & Rust   \\
   & & Brownie \cite{brownie}, Alchemy \cite{alchemy} & Python   \\
   & & Ankr \cite{ankr}, Settlemint\cite{settlemint} & \\
   & & Ocean \cite{oceanprotocol}, Infura \cite{infura}  & \\
   \cmidrule{2-3} 
   & \multirow{3}{*}{\textbf{\makecell[l]{Statistical \\ Tools}}}
   & Deepdao \cite{deepdao}, Snapshot \cite{snapshot}, Tally \cite{tally} & DAO \\
   & & Nansen \cite{nansen}, Token Terminal \cite{tokenterminal}  & Tracing Data Movement\\
   & & Messari \cite{messari}, The Block \cite{theblock}, Web3 Index \cite{web3index}  & Data Dashboard \\
   \cmidrule{2-3} 
   & \multirow{9}{*}{\textbf{Infrastructure}} 
   & Helium \cite{helium} &  Wireless Network   \\
   & & NEON \cite{neon} & EVM in Solana   \\
   & & ENS \cite{ens}, Bonfida \cite{bonfida} & Name Service \\
   & & API3 \cite{api3}, Ankr \cite{ankr}, Pocket Network \cite{pocket} & API \\
   & & Deeper Network \cite{deeper}, Datahub \cite{datahub}  & Gateway \\
   & & Getblock \cite{getblock}  & Node Service  \\
   & & Moralis \cite{moralis}  & SDK  \\ 
   & & Quicknode \cite{quicknode}  & Analytics   \\ 
   & & FigmentLearn \cite{figment}  & Education platform   \\

   \bottomrule
  \end{tabular}
 }
\end{table}

\end{document}